\documentclass{mn2e}
\usepackage{epsfig}

\title[Chemical  analysis of  CH stars - I: atmospheric parameters and elemental abundances]
{Chemical analysis of  CH stars - I: atmospheric parameters and elemental abundances}

\author[Drisya Karinkuzhi et al. ]{Drisya Karinkuzhi$^{1,2}$, Aruna Goswami$^{1}$  \\
    $^{1}$Indian Institute of Astrophysics, Koramangala, Bangalore 560034,
    India; drisya@iiap.res.in, aruna@iiap.res.in\\ $^{2}$ Department of physics, Bangalore university, Jnana Bharathi Campus, Karntaka 560056, India\\
}    
\begin{document}

\date{ Accepted ---;  Received ---;  in original 
form --- \large \bf }

\pagerange{\pageref{firstpage}--\pageref{lastpage}} \pubyear{2013}

\maketitle

\label{firstpage}

\begin{abstract}

Results from  high-resolution spectral analyses of a selected sample of 
CH stars are presented. Detailed chemical composition studies of these objects, which could reveal abundance patterns that in turn provide information regarding  nucleosynthesis and
evolutionary status, are scarce in the literature.
 We conducted  detailed
chemical composition studies  for these objects based on high resolution 
(R ${\sim}$ 42\,000) spectra.  The spectra were  taken from the ELODIE 
archive and  cover the wavelength range from 3900 {\rm \AA}, to 6800 {\rm \AA}, in the wavelength range. 
 We estimated the stellar atmospheric parameters, the effective 
temperature T$_{eff}$, the surface gravity log\,$g$, and metallicity [Fe/H] 
 from Local thermodynamic equilibrium analyses using model atmospheres. Estimated  temperatures 
of these objects cover a wide range from 4550 K to 6030 K, the surface gravity 
from 1.8 to 3.8  and 
metallicity  from $-$0.18 to $-$1.4. We report updates on elemental abundances
for several heavy elements  and present estimates of abundance ratios of 
Sr, Y, Zr, Ba, La, Ce, Pr, Nd, Sm, Eu and Dy with respect 
to Fe. For the object  HD~188650  we present 
the first abundance analyses results based on a high resolution spectrum.
Enhancements of heavy elements relative to Fe, that
are characteristic of CH stars are evident from our analyses for most
of the objects. 
A parametric model based study is performed  to understand the relative
contributions from the s- and r-processes to the abundances of the heavy
elements.
\end{abstract}


\section{Introduction}
 CH stars  characterized by iron deficiency, enhanced carbon and 
s-process elements are  known to be  post-mass-transfer binaries
(McClure \& Woodsworth 1990) in which the  companion  (primary) has 
evolved to white dwarf passing through an AGB stage of evolution. The 
chemical composition of CH stars (secondaries) bear the signature of 
the nucleosynthesis processes occurring in the  companion AGB stars 
due to mass transfer. Two suggested mass transfer mechanisms include  
Roche-lobe overflow and wind accretion. Recent hydrodynamical simulations 
have shown  in the case of the slow  and dense winds, typical of AGB 
stars, that  efficient wind mass transfer is possible through a 
mechanism called wind Roche-lobe overflow (WRLOF) 
(Abate et al. 2013 and references therein).
CH stars (secondaries) thus form ideal targets for  studying 
the operation of s-process occurring in AGB stars. Chemical abundances  
of key elements such as Ba, Eu etc. and their abundance ratios could  
provide insight in this regard.  
However, literature survey shows that detailed chemical composition 
studies of many of the  objects belonging to the CH star catalogue of  
Bartkevicius (1996) are currently not available. A few studies that 
exist are either limited by resolution or  wavelength range. We 
have therefore undertaken to carry out chemical composition studies 
for a selected sample of CH stars from this catalogue using high 
resolution spectra. In our previous studies along this line
we have considered the sample of faint high latitude carbon stars 
of Hamburg/ESO survey (Christlieb et al. 2001) and based on medium
resolution spectroscopy found about 33 per cent of the objects to be potential
CH star candidates (Goswami 2005, Goswami et al. 2007, 2010). Chemical
composition of  two objects from this sample  based on high resolution 
Subaru spectra  are discussed in Goswami et al. 2006.

CH stars (with $-$0.2 $<$[Fe/H] $<$ $-$2) and the class of carbon-enhanced 
metal-poor (CEMP)-s  stars of the CEMP stars classification of Beers \& Chrislieb (2005) are believed to have  a similar origin as far as their 
chemical composition is concerned and that, the CEMP-s stars are thought to be more 
metal-poor counterparts of CH stars. High resolution spectroscopic analyses  
of carbon-enhanced metal-poor (CEMP) stars have established  that 
the largest group of CEMP stars are s-process rich (CEMP-s) stars and 
accounts for about 80 per cent of all CEMP stars (Aoki et al. 2007). 
Chemical composition studies of  carbon-enhanced metal-poor stars 
(Barbuy et al. 2005; Norris et al. 1997a, b, 2002;  
Aoki et al.  2001, 2002a,b; Goswami et al. 2006, 
Goswami \& Aoki 2010) also  have suggested  that a variety of production 
mechanisms are needed to explain the observed range of elemental 
abundance patterns in them; however, the binary scenario of CH star 
formation  is currently considered as also the most likely formation 
mechanism  for CEMP-s stars.  This idea has  gained further 
support with the demonstration by Lucatello et al. (2005b), that 
the fraction of CEMP-s stars with detected radial-velocity variations 
is consistent  with the hypothesis of all being members of binary systems.

Among CEMP stars the group of CEMP-r/s stars show enhancement of both
r- and s-process elements ( 0 $<$[Ba/Eu] $<$ 0.5 (Beers \& Chrislieb 2005)).
Two objects in our sample are found to  show [Ba/Eu] ratios in this range. 
Elements heavier than iron are mainly produced  by two neutron-capture 
processes, the slow neutron-capture process (s-process) and the rapid 
neutron-capture process (r-process). They require entirely different 
astrophysical environments, different time-scales and different neutron fluxes.  While slow neutron-capture elements are believed 
to be produced in the inter pulse phases of low mass AGB stars, the 
rapid neutron-capture process requires very high temperatures and 
neutron flux and are expected to be produced during  supernova explosions. 
To understand the contribution of these two processes to the chemical 
abundance of the neutron-capture elements we have conducted a parametric
model based study. Our study  indicates  three objects in our 
sample to have  abundances of  heavy elements with major contributions
coming from the s-process.

Section 2 describes   the source of our high resolution spectra.
Radial velocity estimates are presented in section 3. Temperature 
estimates from photometry are discussed in section 4. Determination 
of stellar atmospheric parameters are presented  in section 5. Results 
of abundance analysis are discussed in section 6. In section 7 we 
present  brief discussions on each individual star. A discussion on 
the parametric model based analysis is presented in section 8. 
Conclusions are drawn in section 9.

\section{High resolution spectra of the program stars}
Program stars are selected from the CH star catalogue of Bartkevicius (1996);
the basic data of these objects are listed in Table 1. The spectra  
are taken from the  Elodie archive  (Moultaka et al. (2004)).  
We have considered only those CH stars for which high resolution 
spectra are available in the archive with S/N ratio $>$ 20.
ELODIE is an echelle spectrograph used at the 1.93 m telescope of 
Observatoire de  Haute Provence (OHP). Optimal extraction and wavelength 
calibration of data are automatically performed by the online reduction 
software TACOS. The spectra recorded in a single exposure as 
67 orders on a 1K CCD have a  resolution of R ${\sim}$ 42000. The 
wavelength range spans from  3900 {\rm \AA}, to 6800 {\rm \AA}. 
A few  sample spectra are shown in Figures 1 and 2.

\section{Radial velocity}
Radial velocities of our program stars are calculated using  a selected 
set of clean  unblended lines in the spectra. Estimated mean radial 
velocities along with the standard deviation from the mean values 
are presented in Table 2. The literature values  are also 
presented in  for a comparison. Radial velocity variations are
reported in  McClure (1984, 1997) and  McClure \& Woodsworth (1990) 
for four stars, HD 16458, 201626, 216219 and 4395.
Our  radial velocity estimate  for HD 48565  shows a difference 
of $\sim$ 6 km s$^{-1}$ from the literature  value.  Variation   
in radial velocity for this object was also reported by North et al. (1994), 
and Nordstroem et al. (2004). For the remaining stars  we note
a small  difference from the literature values. Except for HD~201626 
($-$141.6 km s$^{-1}$) and HD~81192 ($-$136.5 km s$^{-1}$),  the
program stars are  low-velocity objects.

\begin{figure}
\centering
\includegraphics[height=8cm,width=8cm]{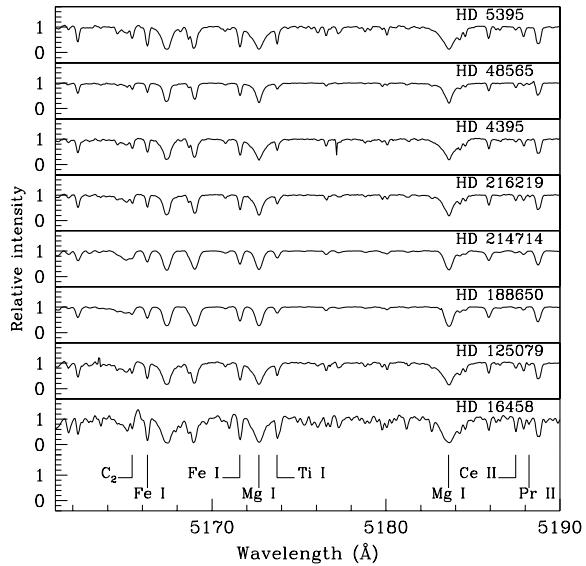}
\caption{Sample spectra of a few program stars in the wavelength region 
5160 to 5190 {\rm \AA}.}
\end{figure}

\begin{figure}
\centering
\includegraphics[height=8cm,width=8cm]{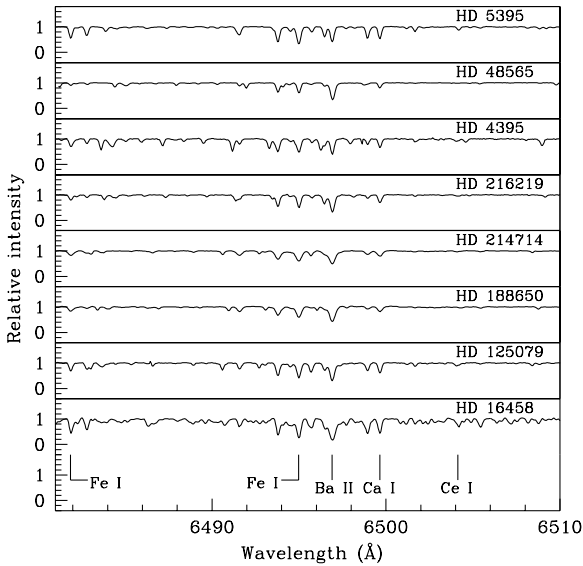}
\caption{ Spectra showing the wavelength region 6480 to 6510 
{\rm \AA}, for the same  stars as  in Figure 1.}
\end{figure}

\begin{table*}
{Table 1: Basic data for the program stars}\\
\begin{tabular}{cccccccc}
\hline
Star Name.   & RA(2000) & DEC(2000)& B& V&J&H&K \\
\hline
HD4395   &	00 46 13.7498&  -11 27 08.567 &	8.39 &	7.70 	&6.394 &6.099 &	5.972 \\
HD5395   &	00 56 39.9051 & +59 10 51.800 &	5.576 &	4.632 &	3.123 &	2.680 &	2.468 \\
HD16458  &	02 47 47.7083&  +81 26 54.512 &	7.122 &	5.790 &	3.840 &	3.343 &	3.032 \\
HD48565  &	06 44 54.9224&  +20 51 38.353 &	7.732 &	7.204 &	6.122 &	5.884 &	5.806 \\
HD81192  &	09 24 45.3354&  +19 47 11.866 &	7.455 &	6.539 &	4.846 &	4.282 	&4.119 \\
HD125079 &	14 17 20.7163&  -04 15 57.819 &	9.57 &	8.67 	&7.136 &6.759 &	6.610 \\
HD188650 &	19 54 48.2509&  +36 59 44.434 &	6.528 &	5.793 &	4.626 &	4.111 &	4.114 \\
HD201626 &	21 09 59.2714&  +26 36 54.928 &	9.20 &	8.13 	&6.315 &5.838 &	5.736 \\
HD214714 &	22 39 34.3326 & +37 35 34.140 &	6.865 &	6.038 &	  -	 & -	      &\\
HD216219 &	22 50 52.1513&  +18 00 07.585 &	8.06 &	7.44 	&6.265 &6.034 &	5.935\\ 

\hline
\end{tabular}
\end{table*}

{\footnotesize
\begin{table*}
{\bf Table 2: Radial velocities}\\
\begin{tabular}{cccc}
\hline
Star Name   & $ V_r$ km s$^{-1}$   &  $ V_{r}$ km s$^{-1}$ &   References \\
            & our estimates       & from literature  &  \\
\hline
HD~4395 &        $-$0.8   $\pm$ 0.8    & $-$1.3    &    4  \\ 
HD~5395    &     $-$47.8  $\pm$ 0.9    & $-$47.0   &    5  \\ 
HD~16458   &    	18.2     $\pm$ 0.5    &    18     &    2 \\ 
HD~48565   &     $-$25.7  $\pm$ 0.4    & $-$19.4   &    4  \\ 
HD~81192  &	136.5    $\pm$ 0.3    &   135.3   &    2 \\ 
HD~125079  & 	$-$4.5   $\pm$ 0.2    & $-$4.3    &    1 \\
HD~188650  &     $-$24.6  $\pm$ 0.4    & $-$23.8   &    2 \\ 
HD~201626 &	$-$141.6 $\pm$ 1.2    & $-$145.7  &    4 \\ 
HD~214714  &     $-$7.0   $\pm$ 0.5    & $-$6.8    &    2 \\
HD~216219  &     $-$6.8   $\pm$ 0.7    & $-$7.5    &    3  \\ 

\hline-
\end{tabular}

1. Smith et al. (1993) 2. Wilson (1953) 3. De Medeiros \& Mayor (1999) 4. Nordstr\"oem et al. (2004)  5. Masseroti et al. (2008)\\
\end{table*}
}

{\footnotesize
\begin{table*}
{\bf Table 3: Temperatures from  photometry }\\
\begin{tabular}{llllllllc}
\hline
Star Name  &  $T_{eff}$ K &  $T_{eff}(-0.5)$ K &  $T_{eff}(-0.5)$ K & $T_{eff}(-1.0)$ K & $T_{eff}(-1.0)$ K & T$_{eff}$(-0.5) K &
T$_{eff}$(-1.0) K & Spectroscopic  \\
           &  (J-K)      &   (J-H)          &   (V-K)          &  (J-H)          &    (V-K) &  (B-V) & (B-V) &estimates (K)\\
\hline
HD~4395    &      5496.71&   5776.28   & 5398.05 & 5791.79  & 5392.69 & 5365.08 & 5243.37  & 5550\\
HD~5395    &    4556.52 &    4950.59  & 4842.04 &  4967.30 &  4829.76 & 4666.20 & 4563.87  & 4860\\ 
HD~16458   &    4082.41 &    4713.82  & 4216.43 &  4730.59 &  4198.88 & 3886.55 & 3804.68  & 4550\\ 
HD~48565   &      6047.46 &    6150.12  & 5891.51 &  6164.60 &  5894.11 & 5928.43 & 5790.39  & 6030\\ 
HD~81192   &	 4321.94 &    4382.20  & 4555.65 &  4398.88 &  4540.63 & 4734.38 & 4630.20  & 4870\\
HD~125079  & 	 5037.98 &    5285.39  & 4964.79 &  5301.79 &  4953.86 & 4774.22 & 4668.96  & 5520\\ 
HD~188650  &     5095.69 &    4553.41  & 5469.31 &  4570.17 &  5464.99 & 5226.76 & 5108.96  & 5700\\
HD~201626  &  	 4829.70 &    4754.07  & 4581.24 & 4770.85  &  4566.44 & 4381.70 & 4286.98  & 5120\\ 
HD~214714  &      --      &     --      &  --     &  --      &   --     & 4964.61 & 4854.13  & 5550 \\ 
HD~216219  &     5969.13 &    6209.69  & 5723.41 &  6223.98 &  5723.10 & 5595.08 & 5466.79  & 5960\\

\hline
\end{tabular}

The numbers in the parenthesis indicate the metallicity values at which 
the temperatures are calculated.\\
\end{table*}
}
\section{ Temperatures from photometric data}

We used colour-temperature calibrations of Alonso, Arribas $\&$ Martinez-Roger (1996) 
that relate $T_{\rm eff}$ with various optical and near-IR colours. Estimated 
uncertainty in the temperature calibrations  is $\sim 100$\,K. The Alonso et 
al. calibrations use Johnson photometric systems for UBVRI and  use TCS 
(Telescopio Carlos Sanchez) system  for IR colours, J-H and J-K. 
The necessary transformations between these photometric systems are 
performed using transformation relations from  Ramirez $\&$ Melendez (2004) 
and Alonso et al. (1996, 1999). The $B-V$ colour of a 
star with strong molecular carbon absorption features depends not only 
on $T_{\rm eff}$, but also on the metallicity of the star and on the strength 
of its molecular carbon absorption features, due to the effect of CH molecular 
absorption in the B band. The derived $T_{\rm eff}$ from V-K is ${\sim}$ 
450 K, and from J-H  is ${\sim}$ 250 K lower than the adopted spectroscopic
 $T_{\rm eff}$   derived by imposing Fe I excitation equilibrium. The 
temperature calibrations  from the  $T_{\rm eff}$ - $(J-H)$ and 
$T_{\rm eff}$ - $(V-K)$ relations  involve a metallicity ([Fe/H]) term. 
Estimates of  $T_{\rm eff}$  at  two assumed  metallicity values 
(shown in parenthesis) are listed in Table 3.

\section{Stellar Atmospheric parameters}  

A set of  Fe I and Fe II lines with excitation potential in the range  
0.0 - 5.0 eV and equivalent width 20 {\rm \AA} to 180 {\rm \AA}  
were selected to find the stellar atmospheric parameters. These lines are 
listed  in Table 4. Throughout our analysis we have assumed 
local thermodynamic equilibrium (LTE). A recent version of MOOG of 
Sneden (1973) is used for the calculations. Model atmospheres were 
selected from the Kurucz grid of model atmospheres with no convective 
over shooting. These are available at http: //cfaku5.cfa.harvard.edu/ 
labelled with a suffix odfnew. Solar abundances are taken from 
Asplund et al. (2005). 

The microturbulent velocity was estimated at a given effective temperature 
by demanding that there be no dependence of the derived Fe I abundance 
on the equivalent width of the corresponding lines. 

The effective temperature is determined by making the slope of the 
abundance versus excitation potential of Fe I lines to be nearly zero. 
The initial value of temperature is taken from the photometric estimates
and  arrived at a final value by an iterative method with the slope 
nearly equal to zero. Figures 3 and 4  show  abundance of Fe I 
and Fe II as a function of  excitation potential and  equivalent 
widths respectively.

The surface gravity was fixed at a value that gives same abundances for 
Fe I and Fe II lines. Clean Fe II lines are more difficult to find 
than Fe I lines, so we were limited to 4-10 Fe II lines for the abundance 
analysis in most cases. 
\begin{figure}
\centering
\includegraphics[height=8cm,width=8cm]{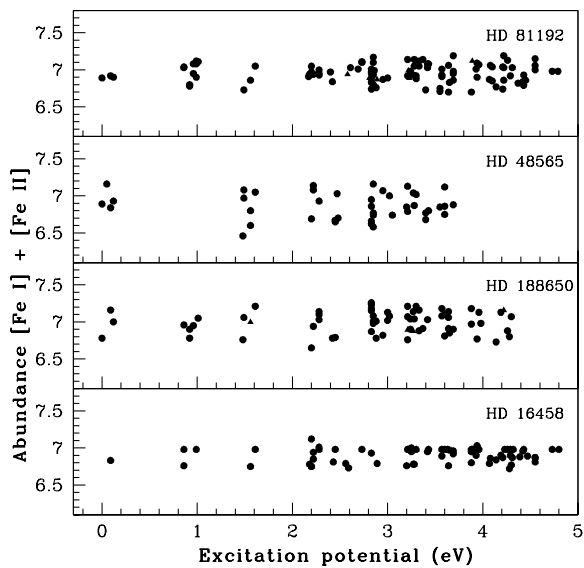}
\caption{The iron abundances of stars are shown for
individual Fe I and Fe II lines as a function of excitation potential.
The solid circles indicate Fe I lines and solid triangles  indicate 
Fe II lines.}
\end{figure}

\begin{figure}
\centering
\includegraphics[height=8cm,width=8cm]{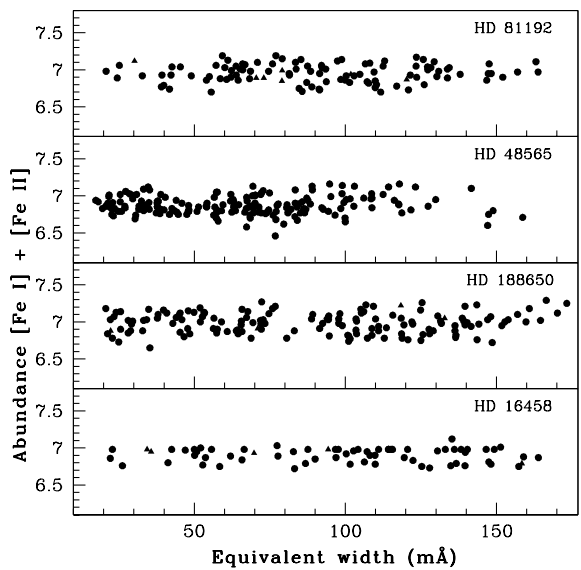}
\caption{The iron abundances of stars  are shown for individual Fe I 
and Fe II lines as a function of equivalent width. The solid circles 
indicate Fe I lines and solid triangles indicate Fe II lines.}
\end{figure}

\begin{table*}\tiny
{\bf  Table 4: Fe lines used for deriving atmospheric parameters}\\
\begin{tabular}{lllcllllllllll}
\hline
Wavelength&    Element    &    E$_{low}$ &   log gf&  HD~4395 &HD~5395&HD~16458&HD~48565&HD~81192&HD~125079&HD~188650&HD~201626&HD~214714&HD~216219\\

\hline
4114.440  & Fe I     &    2.830 &-1.300 &   -       &  -      &  -      &  57.9     &   91.4 & -     &    -       &  -       & 	 115.5  &  -      \\
 4132.900  &      &    2.850 &   -1.010 &   -       &  -	 &  -      &    72.2   &   -      & -     &    -       &  -       & 	 -      &  96.4 \\ 
 4143.870  &      &    1.560 &   -0.510 &   -       &  -	 &  -      &    147.1   &  -       & -     &    -       &  -       & 	 -      &  -     \\ 
 4147.670  &      &    1.490 &   -2.100 &   -       &  -      &  -      &   89.1    &   120.9 & -     &    -       &  -       & 	 -      &  -     \\ 
 4153.900  &       &   3.400 &   -0.320 &   -       &  -	 &  -      &    76.9   &   -      &  -    &    -       &  -       & 	 -      &  -     \\ 
 4154.500  &      &    2.830 &   -0.690 &   -       &  -	 &  -      &    79.6   &   -      &  -    &    -       &  -       & 	 -      &  -     \\ 
 4184.890  &     &     2.830 &   -0.870 &   113.7   &  -	 &  -      &    -       &  -       &  -    &    143.5   &  -       & 	 134.7  &  104.5 \\ 
 4187.040  &       &    2.450 &  -0.550 &   -       &  -	 &  -      &    100.0   &  -       &  -    &    -       &  -       & 	 -      &  137.7 \\ 
 4216.180  &       &    0.000 &  -3.360 &   -       &  -	 &  -      &    88.4   &   -      &  -    &    -       &  -       & 	 -      &  -     \\ 
 4202.030  &      &    1.490  &  -0.710&    -       &  -	 &  -      &    -       &  -       &  -    &    -       &  130.2   & 	 -      &  -     \\ 
 4337.050 &       &    1.560  &  -1.700&    124.3   &  -	 &  -      &    -       &  -       &  -    &    -       &  -       & 	 -      &  125.5 \\ 
 4438.345 &       &    3.882  &  -1.630&    -       &  -	 &  -      &    -       &  -       &  -    &    29.2   &  -       & 	 42.7  &  -     \\ 
 4427.310 &       &    0.050  &  -2.920&    -       &  -	 &  -      &    117.9   &  -       &  -    &    -       &  -       & 	 -      &  -     \\ 
 4422.567 &       &   2.845   &  -1.110&    -       &  -      &  -      &   -        &   123.5 & -     &    -       &  -       & 	 -      &  -     \\ 
 4445.470 &       &    0.087  &  -5.380&    -       &  -	 &  122.4  &    -       &   -      &  -    &    39.8   &  -       & 	 -      &  -   \\   
 4446.833 &       &    3.687  &  -1.330&    -       &  -     &  -      &   34.1    &   77.0  &-      &    -       &  -       & 	 -      &  51.5 \\ 
 4447.720 &       &    2.220  &  -1.340&    -       &  -     &  -      &   98.9    &   138.0  &-      &    141.6   &  38.5   & 	 -      &  -     \\ 
 4454.380 &       &    2.832  &  -1.250&    -       &  -     &  -      &   70.0    &   101.4  &-      &    -       &  -       & 	 -      &  88.7 \\ 
 4461.653 &       &    0.087  &  -3.210&    137.6   &  -	&  -      &    92.3   &    -      & -     &    -       &  -       & 	 -      &  -     \\ 
 4466.551 &       &    2.832  &  -0.590&    -       &  -     &  -      &   100.3    &   147.3  &-      &    160.9   &  91.8   & 	 160.1  &  125.5 \\ 
 4484.220 &       &    3.602  &  -0.720&    81.4   &  -     &  -      &   63.2    &   83.7  &95.3  &    94.7   &  49.7   & 	 -      &  77.7 \\ 
 4489.739 &       &    0.121 &   -3.966&    -       &  -     &  -      &   67.5    &   123.6  &-      &    122.0   &  -       & 	 -      &  -     \\ 
 4531.150 &       &    1.490 &   -2.150&    -       &  -	&  -      &    87.2   &    -      & -     &    -       &  -       & 	 -      &  -     \\ 
 4547.846 &      &     3.546 &   -0.780&    -       &  -     &  -      &   54.2    &   85.7  &96.5     &    -       &  -       & 	 -      &  -     \\ 
 4566.514 &      &     3.301  &  -2.250&    -       &  -	&  55.7  &    -       &   -       & -     &    -       &  -       & 	 -      &  -     \\ 
 4574.215&       &     3.211 &   -2.500&    28.1   &  55.9 &  -      &   -        &    -      &-      &    23.4   &  -       & 	 35.1  &  -     \\ 
 4587.128 &      &     3.573 &   -1.780&    51.3   &  -     &  -      &   -        &   55.0  &-      &    44.9   &  -       & 	 60.7  &  44.5 \\ 
 4595.358 &      &     3.301 &   -1.720&    57.9   &  -     &  -      &   38.9    &   68.3  &-      &    76.7   &  -       & 	 56.1  &  65.9 \\ 
 4596.060 &       &    3.602 &   -1.640&    -       &  -	&  -      &    34.7   &    -      & -     &    -       &  -       & 	 57.2  &  -     \\ 
 4619.287 &       &    3.602 &   -1.120&    -       &  -	&  -      &    41.8   &   -       & -     &    89.2   &  -       & 	 -      &  74.5 \\ 
 4625.044 &       &    3.241  &  -1.340&    -       &  -     &  -      &   -        &   88.5  &-      &    94.0   &  -       & 	 -      &  -     \\ 
 4630.121 &       &    2.278  &  -2.600&    -       &  -     &  -      &   38.1    &   81.5  &-      &    72.8  &  -       & 	 92.2  &  -     \\ 
 4632.911 &       &    1.608  &  -2.913 &   -       &  -	&  -      &    57.6   &  -        & -     &    109.3   &  -       & 	 -      &  -     \\ 
 4635.846 &       &    2.845  &  -2.420&    -       &  -     &  -      &   -        &   63.9  &-      &    42.7   &  -       & 	 65.6  &  33.7 \\ 
 4643.463  &      &    3.654  &  -1.290 &   -       &  82.8 &  -      &   -        &   -       &-      &    57.4   &  -       & 	 73.1  &  52.3 \\ 
 4690.140 &       &    3.686 &   -1.640 &   -       &  67.5 &  -      &   23.0    &   53.9  &-      &    -       &  -       & 	 -      &  44.4 \\ 
 4710.280 &       &    3.018 &   -1.610&    -       &  -	&  -      &    57.3   &   -       & -     &    94.5   &  -       & 	 -      &  -     \\ 
 4733.591 &       &    1.484  &  -2.710&    -       &  -	&  -      &    57.3   &    -      & -     &    -       &  -       & 	 114.0  &  92.5 \\ 
 4736.772 &       &    3.211  &  -0.740&    -       &  -     &  -      &   89.0    &   125.8  &-      &    139.6   &  -       & 	 -      &  -     \\ 
 4741.529 &       &    2.832  &  -2.000&    71.2   &  -     &  -      &   -        &   83.7  &-      &    65.8   &  -       & 	 67.2  &  -     \\ 
 4745.800 &       &    3.654  &  -0.790&    -       &  -     &  -      &   48.2    &   87.2  &-      &    -       &  -       & 	 -      &  -     \\ 
 4768.319 &       &    3.686  &  -1.109&    -       &  -	&  102.7  &    45.9   &   -       & -     &    72.2   &  -       & 	 88.3  &  -     \\ 
 4787.833 &      &     3.000  &  -2.770&    -       &  -	&  -      &    -       &  -        & -     &    24.2   &  -       & 	 30.5  &  -     \\ 
 4788.760 &      &     3.237  &  -1.760&    63.0   &  87.3 &  -      &   31.6    &   -       &-      &    57.6   &  -       & 	 -      &  57.9 \\ 
 4789.650 &      &     3.546  &  -0.910&    87.7   &  -     &  -      &   -        &   84.7  &-      &    -       &  -       & 	 -      &  -     \\ 
 4842.788 &      &     4.104 &   -1.560&    41.1   &  -     &  -      &   -        &   45.4  &-      &    -       &  -       & 	 -      &  -     \\ 
 4871.318 &      &     2.865 &   -0.360&    -       &  -	&  -      &    108.8   &          & -     &    -       &  -       & 	 -      &  145.2 \\ 
 4875.870 &      &     3.332  &  -2.020&    53.2   &  71.4 &  -      &   25.5    &   63.8  &-      &    32.9   &  -       & 	 44.3  &  45.9 \\ 
 4882.140 &      &     3.417 &   -1.640 &   -       &  80.2 &  107.2  &   31.7    &   75.9  &-      &    -       &  -       & 	 63.8  &  51.3 \\ 
 4896.440  &      &    3.884 &   -2.050&    - -     & 	&  -      &    -       &          & -     &    20.7   &  -       & 	 34.9  &  -     \\ 
 4903.310 &       &    2.883 &   -0.930&    -       &  -     &  -      &   84.5    &   144.9  &-      &    139.2   &  -       & 	 139.5  &  110.2 \\ 
 4907.737 &       &    3.430 &   -1.840&    56.7   &  81.5 &  -      &   22.3    &     -     &-      &    -       &  -       & 	 -      &  46.6 \\ 
 4908.029 &       &    4.217 &   -1.285&    33.4   &  -	&  -      &    -       &       -   & -     &    -       &  -       & 	 -      &  -     \\ 
 4917.229 &       &    4.191  &  -1.180 &   58.4   &  55.0 &  -      &   28.5    &     -     &-      &    49.9   &  72.8   & 	 65.6  &  142.4 \\ 
 4924.770 &      &     2.279 &   -2.256&    92.1   &  -     &  151.4  &    54.2   &    -      & -     &    106.1   &  -       & 	 108.6  &  -     \\ 
 4930.315 &      &     3.959 &   -1.350&    -       &  -     & -       &   29.4    &  69.9  &-       &    53.2   &  -       & 	 69.4  &  -    \\  
 4939.690 &       &    0.859 &   -3.340&    -       &  147.5 &  -      &   63.1    &   127.2  &122.1  &    119.4   &  -       & 	 131.7  &  88.5 \\ 
 4967.890  &     &     4.191 &   -0.622&    70.8   &  89.1 &  108.1  &   43.4    &    -      &80.4  &    -       &  -       & 	 73.9  &  63.5 \\ 
 4969.916  &     &     4.216 &   -0.710&    69.5   &  90.2 &  -      &   38.9   &     -     &-      &    -       &  -       & 	 -      &  61.9 \\ 
 4985.250 &      &     3.928 &   -0.560&    91.9   &  110.2 &  109.8  &   57.4   &   93.5  &-      &    -       &  -       & 	 -      &  -     \\ 
 5001.860 &      &     3.881 &    0.090&    -       &  -     &  -      &   77.8    &   111.7  &126.6  &    -       &  65.2   & 	 -      &  -     \\ 
 5005.711 &      &     3.883 &   -0.180&    100.8   &  -     &  -      &    74.1   &   -       & 132.3  &    -       &  -       & 	 -      &  98.2 \\ 
 5006.119 &      &     2.833 &   -0.610&    146.8   &  -     &  -      &   98.5    &   157.0  &-      &    173.3   &  -       & 	 -      &  128.4 \\ 
 5007.728&       &     4.294 &   -1.502 &   -       &  -	 &  -     &    -       &     -     &  50.0   &    -       &  -       & 	 -      &  -     \\ 
 5022.235 &      &     3.984 &   -0.530 &   91.1   &  -	 &  -      &    59.5   &         &  -    &    97.8   &  -       & 	 -      &  -     \\ 
 5028.126  &      &    3.573 &   -1.474&    -       &  -	 &  -      &    -       &         &  -    &    75.9   &  -       & 	 -      &  60.1 \\ 
 5031.915  &      &    4.372 &   -1.670&    -       &  37.2  &  -      &   -        &         & -     &    -       &  -       & 	 24.7  &  -     \\ 
 5044.210  &      &    2.851 &   -2.150&    72.1   &  -      &  -      &   -        &   83.5 & -     &    65.6   &  -       & 	 79.9  &  -     \\ 
 5049.820 &      &     2.279 &   -1.344&    136.0   &  -	 &  -      &    86.5   &    -     &  -    &    157.0   &  -       & 	 161.8  &  115.2 \\ 
 5051.635 &      &     0.915 &   -2.795&    -       &  -	 &  -      &    87.6   &   -      &  -    &    -       &  95.9   & 	 -      &  -     \\ 
 5054.640 &       &    3.639 &   -2.140&    37.1   &  55.7  &  -      &   -        &         & -     &    25.6   &  -       & 	 35.1  &  24.7 \\ 
 5079.224 &       &    2.198 &   -2.067&    -       &  -	 &  -      &    70.1   &         &  -    &    -       &  -       & 	 -      &  -     \\ 
 5083.338 &       &    0.958  &  -2.958&    104.3   &  157.2  &  -      &   74.6    &   147.6 & -     &    136.3   &  -       & 	 -      &  98.6 \\ 
 5088.166  &      &    4.155  &  -1.780&    -       &  -	 &  -      &    -       &         &  -    &    -       &  -       & 	 26.62  &  -    \\  
 5109.650 &        &     4.301 &   -0.980&  59.8   &  -	 &  -      &    35.2   &   -      &  -    &    52.7   &  -       & 	 -      &  -     \\ 
 5127.360 &       &    0.920 &   -3.310 &   94.6   &  148.6  &  -      &   64.7    &   117.0 & 120.4 &    107.3   &  -       & 	 -      &  -     \\ 
 5141.739 &       &    2.424 &   -2.150 &   87.1   &  114.1  &  -      &   48.1    &   93.71 & -     &    80.6   &  -       & 	 102.8  &  76.4 \\ 
 5151.910 &      &     1.011 &   -3.320&    -       &  -	 &  -      &    63.0   &         &  -    &    118.5   &  -       & 	 -      &  -     \\ 
 5159.050 &       &   4.283 &   -0.820 &    52.3   &  79.2  &  83.2  &   32.8    &    -     & -     &    46.6   &  -       & 	 67.1  &  57.6 \\ 
 5166.282&        &   0.000 &   -4.195 &    111.9   &  -      &  -      &   72.0   &   133.8 & -     &    112.9   &  79.1   & 	 129.0  &  -     \\ 
 5171.600 &      &    1.485  &  -1.790 &    156.3   &  -  &  -      &   108.5    &         & -     &    178.4   &  103.1   & 	 164.4  &  135.1 \\ 
 5192.343 &      &    3.000 &   -0.420 &    -       &  -	 &  -      &    108.8   &         &  -    &    164.6   &  -       & 	 163.2  &  -     \\ 
 5187.914 &      &    4.143 &   -1.260 &    53.8   &  -	 &  -      &    19.6   &         &  -    &    25.02   &  -       & 	 38.5  &  -     \\ 
 5194.941 &      &    1.557 &   -2.090 &    119.9   &  170.0  &  -      &   83.3    &   146.8 & -     &    151.8   &  -       & 	 142.7  &  114.4 \\ 
 5195.468 &      &    4.220 &   -0.020 &    89.1   &  111.2  &  -      &   62.0    &    -     & -     &    -       &  -       & 	 -      &  89.1 \\ 
 5198.711 &       &   2.223 &   -2.135 &    98.3   &  -      &  -      &   57.5    &   121.5 & -     &    103.5   &  -       & 	 113.0  &  83.3 \\ 
5215.179 &      &    3.266 &   -0.933 &     -       &  131.9  &  148.2  &   69.8    &   -     & -     &    108.8   &  -       & 	 119.6  &  98.3 \\ 
 5228.403 &      &    4.221  &  -1.290 &    -       &  -      &  -      &   -        &   59.3 & -     &    40.9   &  -       & 	 53.13  &  -     \\ 
 5226.862 &      &    3.038  &  -0.667 &    -       &  -	 &  -      &    103.0   &         &  -    &    -       &  -       & 	 -      &  -     \\ 
 5232.939 &      &    2.940  &  -0.060 &    -   &  -	 &  -      &    127.4   &   -      &  -    &    -       &  -       & 	 -      &  167.4 \\ 
 5250.645 &       &   2.198  &  -2.050 &    -       &  -	 &  -      &    -       &         &  -    &    113.6   &  -       & 	 130.3  &  74.9 \\ 
 5247.050 &      &    0.087  &  -4.980&     -       &  -      &  -      &   22.9    &   100.3 & -     &    -       &  -       & 	 -      &  -     \\ 
 5242.490 &      &    3.630 &   -0.970&     74.2   &  -	 &  111.0  &    50.9   &    -     &  -    &    -       &  -       & 	 90.8  &  -     \\ 
 5253.461&       &    3.283 &   -1.670&     73.8   &  -      &  109.9  &   35.2    &   85.4 & -     &    60.9   &  -       & 	 73.4  &  61.7 \\ 
 5263.305 &      &    3.265 &   -0.970&     105.5   &  -	 &  -      &    68.5   &         &  -    &    101.1   &  67.5   & 	 103.2  &  93.9 \\ 
 5266.555 &      &    2.998 &   -0.490 &    106.1   & 	 &  -      &    -       &         &  -    &    -       &  152.2   & 	 127.2  &  -     \\ 
 5281.800 &      &    3.040 &   -0.833&     128.0   &  -      &  -      &   81.9    &   147.2 & -     &    125.0   &  76.3   & 	 131.9  &  112.1 \\ 
 5283.630 &      &    3.240 &   -0.524 &    -       &  -	 &  -      &    95.3   &         &  -    &    147.1   &  -       & 	 148.1  &  -     \\ 
 5324.178  &     &    3.211 &   -0.240&     -       &  -      &  -      &   112.5    &   172.4 & -     &    170.2   &  98.85   & 	 169.2  &  -     \\ 
 5307.370  &     &    1.610  &  -2.912&     85.3   &  130.0  & 149.3   &   47.5    &   112.5 & 107.6 &    99.0   &  57.19   &      104.6 &  73.1 \\  
\hline
\end{tabular}
\end{table*}

\begin{table*}\tiny
{\bf  Table 4 : continued }\\
\begin{tabular}{lllcllllllllll}
\hline
Wavelength&    Element    &    E$_{low}$ &   log gf&  HD~4395 &HD~5395&HD~16458&HD~48565&HD~81192&HD~125079&HD~188650&HD~201626&HD~214714&HD~216219\\
\hline
 
 5321.110 &      &    4.434  &  -1.090&     39.3   &   53.7    &      -      &  -      &   39.9   &    -     &   22.9 & - &  35.1  &  30.3\\
 5322.040 &      &    2.279  &  -2.800&     55.8   &   95.7    & 114.3     &  -      &   81.4   &   75.8  &   48.5 & -  &  63.0  &  -      \\      
 5328.040 &      &    0.915   & -1.470&     -       &   -        &	  -  &  147.4  &  	 &   -      &   -     & 160.5	&  -      &  -      \\      
 5339.928&        &   3.266  &  -0.680&     124.6   &   148.2    &	  -   &  83.38  &   121.5   &   -  &   129.1 & 78.3	&  134.2  &  108.2  \\      
 5364.858 &      &    4.446  &   0.220&     107.0   &   120.8    &	  -        &  71.3  &   100.0   &   119.1  &   111.0 & - &  109.2  &  99.9\\      
 5367.479 &       &   4.415  &   0.350 &    -       &   123.6    &130.5    &  70.1  &   107.2   &   -      &   118.6 & 63.4	  &  115.8  &  104.4\\      
 5369.958 &       &   4.371  &   0.350&     124.2   &   136.2    &	  -        &  84.0  &   107.7   &   - &   131.5 & 66.5&  134.9  &  109.6  \\      
 5373.698 &       &   4.473  &  -0.860&     -       &   72.0    &	  77.6	   &  23.6 &    -   	 &   -      &   -   & - &  48.9  &  47.4  \\      
 5379.574 &      &    3.694  &  -1.480&     -       &   81.6    &  100.3    &  23.2  &   68.6   &   67.0  &   45.3 & - &  54.5  &  48.8 \\      
 5383.369 &      &    4.312  &   0.500&     132.6   &   148.6    & 163.9 &  92.7  &   134.4   &   143.5  &   136.3 & -     &  139.4  &  125.2  \\      
 5429.710 &      &    0.960 &   -1.881&     -       &   -      &	  -   &  141.7  &  	 &   -   &   -  & 140.4	  &  -      &  -      \\      
 5434.520 &       &   1.011  &  -2.130&     159.5   &   -        &	  -        &  102.8  &  -	 &   -      &   -     & 128.4 & - &  132.5  \\      
 5464.280 &      &    4.143 &   -1.400&     38.6   &   53.9    &	  -    	   &  18.2  &     -  	 &   55.0  &   -     & -   &  -  &  36.9  \\      
 5497.520 &      &    1.011 &   -2.830&     -       &   -        &	  -   &  -      &   163.1   &   -      &   166.5 & -    &  -      &  -      \\      
 5501.464 &      &    0.958  &  -2.950 &    116.8   &   -   &	  -   &  77.1  &   148.3   &   151.0  &   144.2 & -    	  &  156.8  &  106.7 \\	      
 5525.539  &     &    4.231 &   -1.330&     45.3   &   66.6    & 50.0    &  -      &   -       &   46.3 &   33.2 & -    	 &   37.4  &  37.3  \\      
 5543.937 &       &   4.217 &   -1.140&     56.3  &   75.0    &	  -        &  25.5  &   60.1   &   66.2  &   45.9 & -   &   -      &  44.9  \\      
 5554.882 &       &   4.548  &  -0.440 &    -       &   -        & 97.0    &  -      &   79.1   &   -      &   69.4 & -   &   -    &  -      \\      
 5560.207 &      &    4.434 &   -1.190 &    41.5   &   -        & 42.5    &  17.4  &   42.3   &   49.0  &   29.6 & -    &   40.4  &  35.93  \\      
 5567.392 &      &    2.608&    -2.800&     -       &   -        &	  -        &  21.7  &   81.2   &   -      &   51.9 & -  &   68.2  &  49.9  \\      
 5569.618 &      &    3.417 &   -0.540&     124.3   &   -        &	  -      &  85.7  &   128.3   &   -     &   127.9 & -  &   124.5  &  112.0  \\      
 5576.090 &      &    3.430  &  -1.000&     102.3   &   -        & 133.1    &  67.3  &   106.6   &   111.5  &   110.0 & -    &   -      &  -      \\      
 5586.756 &      &    3.364  &  -0.210&     -       &   -        & -   &  114.3  &  	 &   174.9  &   159.8 & 90.70	 &   153.2  & 138.2  \\      
 5615.660 &   &     3.330 &    0.050 &      -       &   -        &	  -   &  117.7  &  	 &   -  &   -     & -    	 &   -    &  152.1  \\      
 5617.186 &      &    3.252 &   -2.880 &    -       &   41.3    &52.1	   &  -      &       	 &   -      &   -     & -    	 &   -    &  -      \\      
 5618.630 &      &    4.209 &   -1.270 &    43.2  &   61.7    &	  -     &  -      &   41.8   &   -      &   28.1& -    	 &   36.9  &  32.28  \\      
 5636.694 &      &    3.640 &   -2.610&     -       &   34.0   &	  26.3    &  -      &   25.2   &   21.5  &   -     & -  &   -    &  -      \\      
 5701.549&    &     2.559 &   -2.216&       -       &   -     &	  136.7    &  42.6  &  	 &   -      &   -     & -    	 &   -      &  -      \\      
 5731.765 &   &     4.256 &   -1.300&       52.2   &   -        &	  -        &  21.8  &  	 &   -      &   38.0 & -    	 &   46.7  &  -     \\      
 5741.848 &      &    4.256 &   -1.730&     22.5   &   45.3    &	  -    	   &  -      &       	 &   -      &   -     & -    &  - &  -      \\      
 5753.121&    &     4.260 &   -0.760 &      75.2   &   90.2    &87.6 &  37.8  &       	 &   79.5  &   65.5 & -    	 &   73.7  &  65.23  \\      
 5809.220 &      &    3.884 &   -1.610 &    45.8   &         &	  82.8    &  -      &   70.1    &   60.2  &   -     & 18.9	 &   36.9  &  33.2  \\      
 5855.080 &       &   4.607 &   -1.480 &    25.3   &   -        &	  -    &  -      & -&   24.7  &   -     & -    	 &   -      &  -      \\      
 5856.100 &      &    4.294 &   -1.560 &    28.7   &   -        &	  66.5    &  -      &   32.7   &   40.1  &   -     & -    &   -    &  20.1  \\      
 5862.370 &      &    4.550 &   -0.250 &    79.4   &   93.3    &106.3	   &  44.7  &       	 &   86.2  &   68.7 & -    	 &   75.3  &  70.5  \\      
 5883.813 &      &    3.959&    -1.360&     57.8   &   81.4    &	  104.2    &  30.9  &   66.8   &   -   &   -     & -    &   -      &  49.1  \\      
 5914.194 &       &   4.607 &   -0.050 &    -   &   -  &	  -   &  -      &   -       &   -      &   105.1 & -     &   116.2  &  -      \\      
 5976.779 &      &    3.943&    -1.310&     -       &   106.6    &	  -        &  22.7  &   62.1   &   -      &   -     & 21.2&   -      & 51.1 \\      
 6003.017 &      &    3.881 &   -1.120 &    -       &   82.0    &	  97.7	   &  37.8  &       	 &   88.6  &   71.9 & 24.0&   72.9&  67.8  \\      
 6024.049 &      &    4.548 &   -0.102&     61.7   &   -        &	  119.5    &  56.8  &   92.1   &   107.3  &   97.9 & 39.7&   98.0  &  81.1  \\      
 6082.710 &      &    2.223 &   -3.550 &    -       &   -        &	  90.0    &  35.6  &   76.0   &   50.8  &   -   &      	 &   -    &  31.0  \\      
 6136.615 &      &    2.453 &   -1.400&     137.6   &   -        &	  -   &  86.2  &  	 &   -      &   -     & -    &   -      &  -      \\      
 6136.993 &       &   2.198 &   -2.950&     75.5   &   -      &	  125.3    &  26.9  &  	 &   150.6  &   -     & -    	 &   -      &  -      \\      
 6137.694 &       &   2.588 &   -1.400&     -       &   -     &	  -        &  80.6  &  	 &   -      &   139.6 & 83.2	 &   145.9  &  114.6  \\      
 6151.620 &      &    2.176&    -3.370&     32.4   &   -        &	  101.6    &  52.2  &   -    &   64.3  &   34.2 &      	 &   -      &  49.3 \\      
 6165.360 &      &    4.143&    -1.460 &    44.2   &   60.8   &	  65.7    &  -      &   39.1   &   51.6  &   25.5 & -   &   32.1  &  28.6  \\      
 6180.205&       &    2.727 &   -2.780&     50.5   &   -        &	  115.6    &  -      &   70.9   &   -      &   37.8 & -    &   - &  40.1  \\      
 6173.343&       &    2.223 &   -2.880&     67.7   &   107.5    &	  139.7	   &  25.9  &       	 &   85.6  &   67.3 & -    &   -  &  54.2  \\      
 6219.280 &      &    2.197 &   -2.430 &    90.2   &   132.1    & 157.4  &  51.4  &   109.5   &   109.0  &   103.5 & 66.9&   105.6  &  80.2  \\      
 6232.639 &       &   3.653 &   -1.270 &    78.0   &   103.1    &	  -  &  37.4  &       -&   -      &   64.5 & -   &   80.2 &  68.5 	\\      
 6230.736 &       &   2.559 &   -1.280 &    140.6   &   -     &	  -        &  86.9  &  	 &   -      &   153.9 & -    	 &   156.6  &  121.5  \\   
 6246.327 &       &   3.602  &  -0.960&     -       &   -        &	  -   &  62.9  &  	 &   -  &   103.7 & -    	 &   107.1  &  -      \\  
 6240.650 &      &    2.223 &   -3.170&     -       &   87.2 &	  -    &  -      &       	 &   64.9  &   -     & -    	 &   -      &  34.0 \\      
 6252.550  &     &    2.404  &  -1.690&     104.0   &   77.4    &	  -   &  114.9  &   158.0  &   134.1  &   125.7 &  -&   81.8  &  131.7 \\      
 6254.250 &      &    2.278 &   -2.400&     -       &   -        & -  &  -      &  	 &   -      &   -     & -    	 &   122.4  &  -      \\      
 6265.130 &      &    2.176 &   -2.540 &    -       &   -      &- 44.18   &  104.8    &  103.3   &  92.4 & -    	 &   105.9  &  73.3 \\       
 6246.327 &      &    3.602 &   -0.960 &    -       &   -        &120.6    &  -   &  -&   -      &   -     & -    	 &   -      &  -      \\      
 6270.240 &      &    2.858 &   -2.710 &    49.2   &   -        &	  -   &  -      &   62.3   &   -   &   -     & -    	 &   -    &  36.9  \\      
 6280.630  &     &    0.859 &   -4.390 &    -       &   118.2    &	  140.4	   &  -      &      -&   89.4  &   -     & -   &   -   &  -      \\      
 6297.800 &      &    2.222 &   -2.740&     -       &   -        &	  -   &  33.2  & -&   86.7  &   73.4 & -    	 &   82.6  &  68.8  \\      
 6301.498 &      &    3.653  &  -0.740&     99.6   &   -        &	  -   &  63.6  & -&   118.1  &   101.1 & -    	 &   -      &  91.1  \\      
 6318.018 &      &    2.453  &  -2.330&     -       &   -        & 138.3    &  -  &   113.4   &   -   &   106.8 & -    	 &   117.2  &  -      \\      
 6322.690 &       &   2.588  &  -2.402 &    79.4   &   109.3    &	  127.9	   &  37.74  &    -	 &   84.7  &   63.9 & 43.0&   71.0  &  61.3 \\      
 6335.340 &       &   2.197 &   -2.230&     96.6   &   141.4    &-   &  56.4   &  123.4    &  -       &  108.8 & -    	 &   114.4  &  83.96  \\      
 6336.823 &       &   3.686 &   -1.050&     -    &   -    &	  -        &  56.6  &  	 &   -      &   98.2 & -    	  &  99.2  &  -      \\      
6408.016  &     &    3.686 &   -1.040 &     90.3&   -        &	  -    &  48.0  &  	 &   -      &   -     & -  &  100.3  &  76.1  \\      
 6411.650 &      &    3.653 &   -0.820&     -    &   -        &	  -    &  70.3  &   91.5   &   -      &   -     & -    	  &  117.2  &  -      \\      
 6419.980  &     &    4.733 &   -0.240&     79.7   &   96.3    &	  114.3    &  38.6  &   74.6   &   86.1  &   72.1 & -   &  76.8  &  68.8  \\      
 6421.350  &     &    2.278 &   -2.030 &    -       &   -        &	  -        &  68.3 &   127.3   &   132.6  &   122.3 & 75.69&  128.9  &100.0 \\      
 6481.869  &     &    2.278  &  -2.984 &     -       &   105.2    &	  -    	   &  33.2  &       	-&   75.5  &   65.8 & -   &  68.1  &  48.2  \\      
 6494.980 &       &   2.404  &  -1.273 &    -    &   -        &	  -  &  94.4 &   163.8   &   154.1  &   148.6 & -    	  &  166.1  &127.7  \\              
 6574.240  &     &    0.990 &   -5.040 &    31.7   &   -    &	  115.1    &  -      &   66.0   &   -      &   -     & -    &  -      &  -     \\       
 6575.022  &     &    2.588 &   -2.820 &    -    &      -&	  -     &  -      &  	 &   -      &   -     & -       &  -      &  44.39  \\ 	           
 6593.880  &     &    2.432 &   -2.420&     -       &   -  &	  147.5    &  40.02  & -&   -      &   -     & -    	  &  -      &  73.41  \\      
 6597.557 &      &    4.796 &   -1.070&     -       &   -  &	  22.9    &  -      &   45.0   &   40.6  &   -  & -    	  &  -      &  -      \\      
 6627.540  &     &    4.548 &   -1.680&     22.9   &   -        &	  -        &  -      &  	 &   -  &   -     & -   &  -      &  -      \\      
 6677.989 &      &    2.692 &   -1.470&     -       &   -        &	  -        &  -      &   133.8   &   -  &   138.7 & -   &  -      &  107.4  \\      
 6713.750 &      &    4.795 &   -1.390&     -       &   27.0    &	  -  &  -   &   20.8   &   -      &   -     & -    	  &  -    &  -      \\      
 6725.360 &      &    4.103 &   -2.170&     -       &   27.8    &	  -    	   &  -      &       	 &   23.7  &   - & -   &  -      &  -      \\      
 6739.520 &      &    1.557 &   -4.790&     -       &   -        &	  58.4    &  -      &  	 &   23.7  &   -     & -    &  -      &  -      \\      
 6750.150 &      &    2.424 &   -2.620&     -       &   113.5    &	  -        &  33.9  &   74.2   &   81.5 &   -     & -   &  -      &  61.1  \\      
 6752.711 &      &    4.640 &   -1.200&     -       &   -        &  -   &  -      &  	- &   34.0  &   -     & -    	  &  27.0  &  23.4  \\      
 6793.260  &     &    4.076 &   -2.330 &    -       &   21.2    &	  22.3	   &  -      &      -&   -      &   -     & -    	  &  - &  - \\      
4233.172  &  Fe II &   2.580 &   -2.000&    -       &   -        &-  &  97.5  &   101.8   &   -      &   -     & 76.9	  &  -      &  -      \\      
 4491.405  &       &    2.855 &   -2.700&   69.1   &   -        &	  -        &  64.6  &  	 &   -      &   132.9 & 45.7	  &  109.5  &  89.4 \\      
 4515.339 &        &    2.840 &   -2.480&   -    &   -        &	  -        &  74.3  &  	 &   -      &   -     & -    	  &  -      &  101.5  \\      
 4520.224  &       &    2.810 &   -2.600&   -       &   -        &	  - &  71.7  &   70.5   &   -      &   -     & -    	  &  -  &  94.9  \\      
 4620.510 &         &   2.830  &  -3.290&   -       &   66.4    &	  69.8	   &  -      &   -&  -  &   -     & -    	  &  -      &  -    \\      
 4629.340 &         &   2.807  &  -2.280&   -   &   -        &	  -        &  75.8  &  	 &   -      &   99.4    & -    	  &  130.6  &  -      \\          
 4923.930 &         &   2.891  &  -1.320&   145.6   &   161.0    &	  158.4    &  123.6  &   120.3   &   - &   -     & -   &  -  &  165.0\\	      
 5197.577  &        &   3.230  &  -2.100&   -       &   95.0    &	  -  &  71.2  &   73.0   &   -      &   -     & 54.9	  &  131.1  &  98.1\\      
 5534.834  &       &    3.245  &  -2.930&   -       &   -     &	  -  &  50.6  &  	 &   -      &   118.4 & -    	  &  100.4  &  72.9 \\     
 6247.550  &       &    3.891  &  -2.510&   38.3   &   -     &	  34.4    &  42.5  &  -&   -      &   -     & -    	  &  90.7  &  -      \\      
 6369.460  &       &    2.891  &  -4.020&   -       &   26.8    &	  -  &   -    &     - &   -      &   -     & -    &  -      &  -     \\       
 6416.919  &       &    3.891  &  -2.740&    -      &   46.0    &	  35.7  &    25.7  &30.2  &    44.0  &   71.0 & -   &  57.9  &  49.3\\	      
 6456.383  &        &   3.903  &  -2.075&    57.9  &   67.8    &	  -  &    51.4  &	 -      &    -      &   -    &  33.4&  - &  -    \\

\hline
\end{tabular}
\end{table*}

\begin{table*}
{\bf  Table 5: Derived atmospheric parameters for the program stars}\\
\begin{tabular}{lccccc}
\hline
Star Name.   & $T_{eff}$  &  log g  &  $\zeta $ &[Fe I/H] &[Fe II/H]    \\
             &   K        &         &  km s$^{-1}$ &      &             \\
\hline
HD~4395 &      5550    & 3.66   & 0.93	 &  -0.16    & -0.19    \\
HD~5395    &   4860   & 2.51 & 1.21	 &    -0.24 &   -0.24      \\ 
HD~16458   &   4550 &  1.8& 1.92	 &   -0.65   &  -0.66     \\ 
HD~48565   &6030 &3.8&     1.13   & -0.59& -0.59       \\ 
HD~81192  &4870	& 2.75  & 1.08	 & -0.50    & -0.51         \\
HD~125079  & 5520 & 3.3 &1.25	 & -0.18 & -0.18       \\ 
HD~188650  &   5700   & 2.15	&2.33	 &  -0.46 & -0.44     \\ 
HD~201626 &	5120& 2.25  &	1.02  & -1.39    &  -1.41        \\ 
HD~214714  &   5550    &  2.41   &   1.96   &  -0.35    &    -0.36       \\
HD~216219  &  5950   &  3.5 & 1.31 &  -0.17    &   -0.18 \\

\hline
\end{tabular}
\end{table*}

\section{Abundance Analysis}
 Abundances for most of the elements  are  determined  from  the measured  
equivalent widths of lines due to neutral and ionized elements using a recent 
version of MOOG of Sneden (1973) and  the adopted model atmospheres. 
A master line list including all the elements is generated comparing the 
spectra of the program stars with the spectrum of Arcturus. The 
presented line lists contain only those lines which are used for  
abundance calculation. In most of the spectra we could get very few usable 
clean lines. 
The log gf values for the atomic lines were adopted from various sources 
which include  Aoki et al. (2005, 2007), Goswami et al. (2006), 
Jonsell et al. (2006), Luck \& Bond (1991) and Sneden et al. (1996), 
whenever available, and also from Kurucz  atomic line database 
 (Kurucz 1995a,b).
The log gf values for a few La lines are taken from
Lawler, Bonvallet $\&$ Sneden (2001). 
We determined  abundances for  Na, Mg, Ca, Sc, Ti, V, Cr, Mn, Co, Ni, Zn
and for heavy elements Sr, Y, Zr, Ba, La, Ce, Pr, Nd, Sm, Eu and Dy. 
For the elements Sc, V, Mn, Ba, La and Eu, spectrum synthesis
 was used to find the abundances considering  hyperfine 
structure. 
The line lists for each region that is synthesized are taken from Kurucz atomic
line list (http://www.cfa.harvard.edu/amdata/ampdata/\\kurucz23/sekur.html). As an  example,   spectrum synthesis calculations for 
Sc, Ba and La are shown  in  Figures  5, 6 and 7.

 Derived abundance ratios with respect to iron are listed in Table 6. 
In Table 7, we have presented [ls/Fe], [hs/Fe] and [hs/ls] values,
where ls represents light s-process elements Sr, Y and Zr and hs
represents heavy s-process elements Ba, La, Ce, Nd and Sm. Lines used 
for the abundance calculation of these elements are listed in 
Tables 8 and 9. 
\begin{figure}
\centering
\includegraphics[height=8cm,width=8cm]{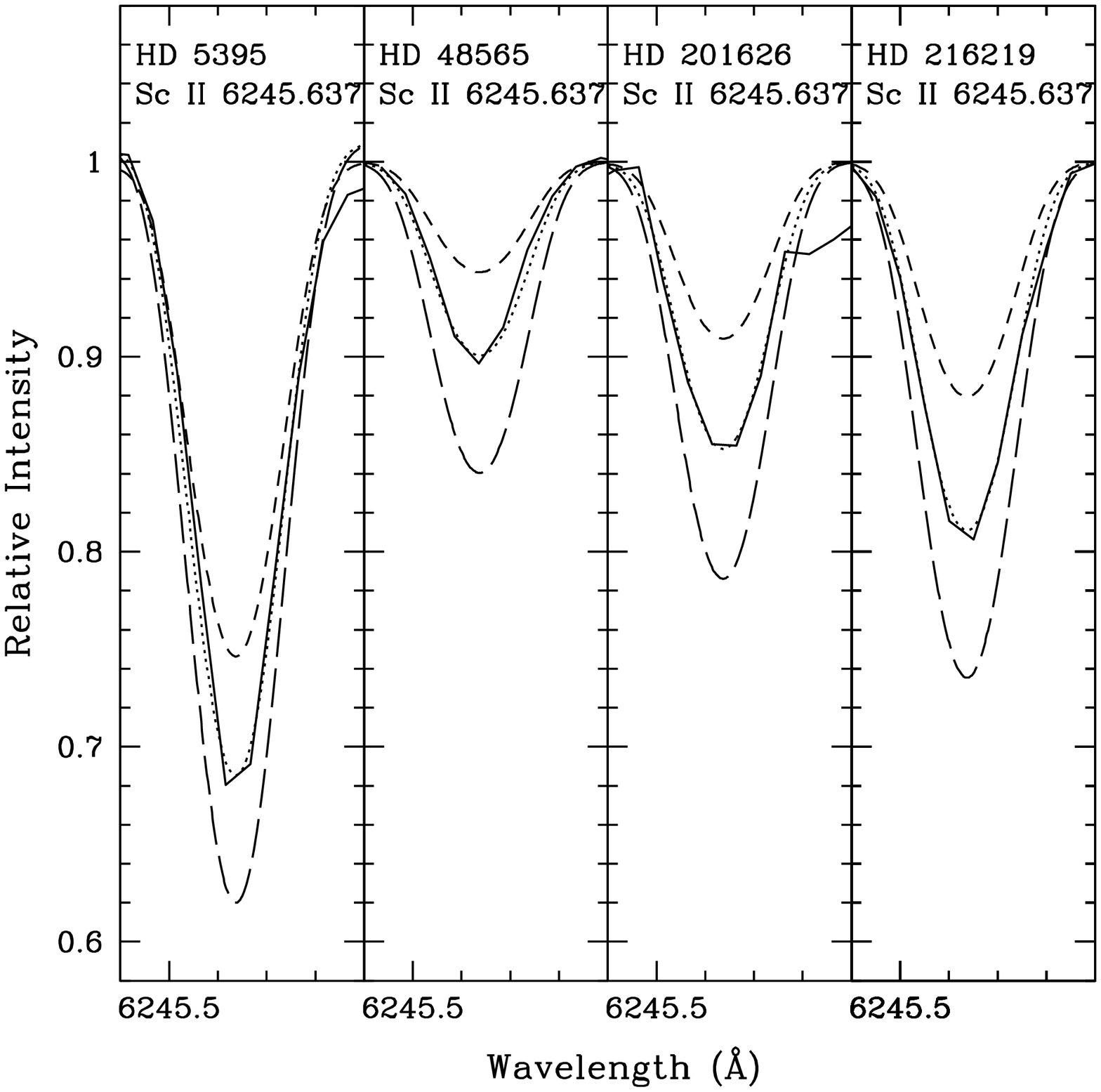}
\caption{ Spectral-synthesis fits of Sc II line at 6245.637 {\rm \AA}. The 
dotted lines indicate the synthesized spectra and the solid
lines indicate the observed line profiles. Two alternative synthetic spectra
for [X/Fe] = +0.3 (long-dashed line) and [X/Fe] = -0.3 (short-dashed
line) are shown to demonstrate the sensitivity of the line strength to the
abundances.}
\end{figure}

\begin{figure}
\centering
\includegraphics[height=8cm,width=8cm]{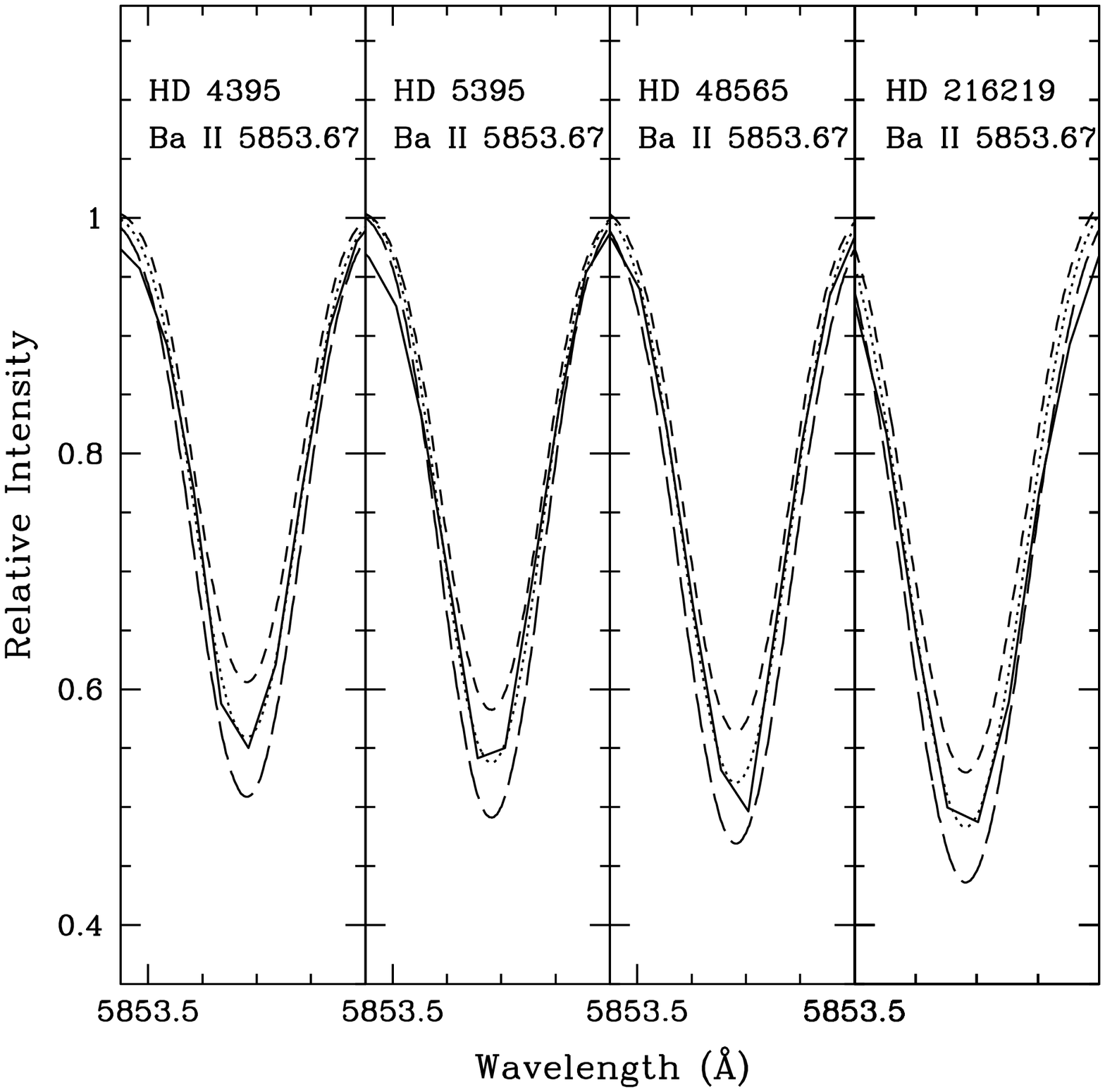}
\caption{ Spectral-synthesis fits of Ba II line at 5853.67 {\rm \AA}. 
The dotted lines indicate the synthesized spectra and the solid
lines indicate the observed line profiles. Two alternative synthetic spectra
for [X/Fe] = +0.3 (long-dashed line) and [X/Fe] = -0.3 (short-dashed
line) are shown to demonstrate the sensitivity of the line strength to the
abundances.}
\end{figure}

\begin{figure}
\centering
\includegraphics[height=8cm,width=8cm]{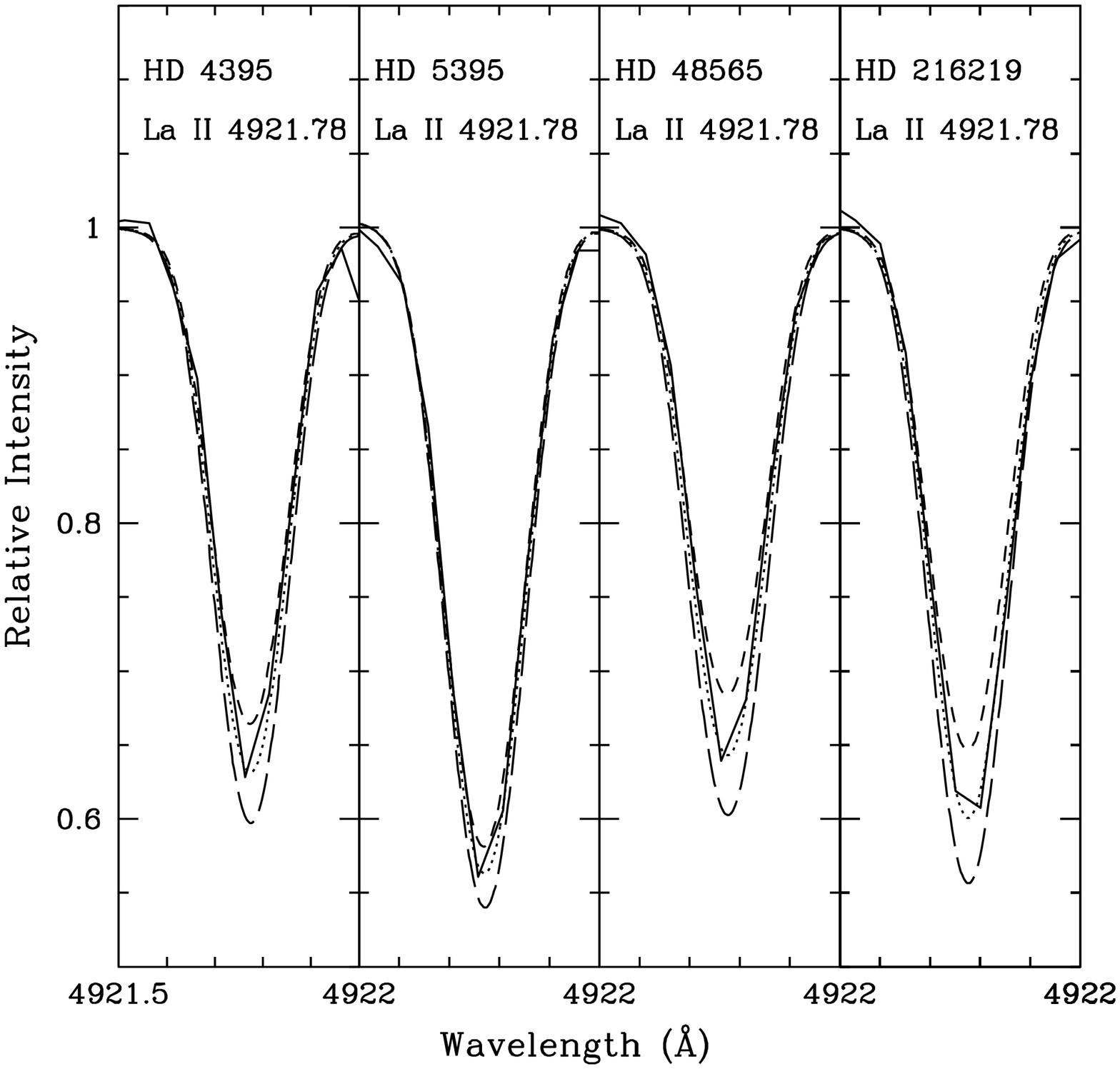}
\caption{ Spectral-synthesis fits of La II line at 4921.78 {\rm \AA}. 
The dotted lines indicate the synthesized spectra and the solid
lines indicate the observed line profiles. Two alternative synthetic spectra
for [X/Fe] = +0.3 (long-dashed line) and [X/Fe] = -0.3 (short-dashed
line) are shown to demonstrate the sensitivity of the line strength to the
abundances.}
\end{figure}

\subsection{Carbon}
Smith (1984) reported a carbon abundance 
of 8.70 dex  and C$^{12}$/C$^{13}$ ${\sim}$ 15 for HD 16458.  For the 
three sub-giant CH stars in our sample, HD 125079, 216219 and 4395 
carbon abundances and isotopic ratios are available in the literature 
(Smith et al. (1993) and Luck \& Bond (1982)). Luck \& Bond (1982) 
determined a [C/Fe] value of 0.4 and 1.2 for  HD 216219 and 4395 
respectively. Smith et al. (1993) gave carbon abundances of  8.65 dex, 9.05 dex,  and 9.02 dex for  HD~4395, 125079 
and 216219 respectively. Vanture (1992a,b) reported a carbon 
abundance of 8.4 dex and  C$^{12}$/C$^{13}$ ${\sim}$ 25 for the 
object HD~201626. Luck (1991) gives a [C/H] ${\sim}$ $-$0.16  and 
C$^{12}$/C$^{13}$ ${\sim}$ 25 for  HD~214714. Baird et al. (1975) 
also noticed an enhancement of carbon in HD~214714 and 188650  
with respect to $\beta$ Aqr. Because of the carbon deficiency 
in $\beta$ Aqr they concluded that these two stars show normal carbon abundances. For HD~81192, Cottrell and Sneden (1986) 
reported a C/N ratio of 11.2 and Shetrone, Sneden \& Pilachowski (1993) gave 
C$^{12}$/C$^{13}$ ${\sim}$ 35. 

\subsection{Na and Al}
The sodium (Na) abundances were calculated using  the lines at 
5682.65 {\rm \AA} and 5688.22  {\rm \AA}  in the case of 
six objects. For HD~48565 a single line at 5682.65 {\rm \AA} is used. 
As these  lines could not be used for HD~5395 and HD~81192, the resonant 
doublet lines at 5890.9 and 5895.9 {\rm \AA} are used. For 
HD~201626 the line at 5895.9 {\rm \AA} is used which is observed 
as a broad line with an equivalent width of 214 m{\rm \AA}. We 
have used LTE analysis for the abundance determination. However, the 
resonance lines are sensitive to non-LTE effects (Baum\"uller \& Gehren 1997;
Baum\"uller, Butler $\&$ Gehren 1998; Cayrel et al. 2004). Derived Na 
abundances from LTE analysis range from $-$0.23 to 0.76 in the present 
sample. We note that  Na in  HD~5395 and HD~81192 is found to be 
underabundant with respect to Fe while the other stars  show mild over 
abundances.
 
Al lines in our spectral  region are blended and could not be used 
for abundance determination.

\subsection{ Mg, Si, Ca, Sc, Ti, V}
We could measure several lines due to these elements. Except for 
HD~201626, that shows an overabundance of Mg with [Mg/Fe] ${\sim}$ 0.69,  
 all other stars  show  mild enhancement or near-solar value of Mg.
 None of the Si lines detected in our spectra was found to be  
usable for abundance determination. Ca  shows a  mild enhancement or 
near-solar value in these objects except for HD~201626, which shows an overabundance of Ca 
with [Ca/Fe] ${\sim}$ 0.65.

Sc abundance is determined using spectrum synthesis calculation of 
Sc II line at 6245.63 {\rm \AA} considering hyperfine structure from 
{\bf Prochaska} and Mc William (2000). While Sc is found to be mildly underabundant
in HD~188650 and HD~214714 with [Sc II/Fe] ${\sim}$ $-$0.5, the ther stars show 
nearly solar values.

Except for HD~201626,  the  program stars show a mild overabundance
or near-solar abundance for Ti measured using more than ten lines. HD~201626 shows an overabundance of Ti with [Ti/Fe] ${\sim}$ 0.8.

We detected more than  16 V I lines but  only one or two 
usable lines of V I  for the determination of V abundance; other lines 
appear either blended or distorted in the spectra. Abundance of V is 
estimated from spectrum synthesis calculation of V I line at 
5727.048 ${\rm \AA}$  taking into account the hyperfine components 
from Kurucz database for all the program stars except HD 201626. 
V abundance could not be estimated for HD~201626 due to severe line 
blending. While HD~16458 shows [V/Fe] ${\sim}$ 0.25 all other program 
stars show  mild under abundance or nearly solar values for V.

\subsection{Cr, Co, Mn, Ni, Zn}
 HD~16458  shows a mild overabundance of Cr  relative to
Fe. HD~125079, 188650, 216219, and 4395 show near-solar 
abundances. The rest of the stars in our sample are mildly underabundant 
in Cr.  HD~201629 however shows a larger  underabundance with 
[Cr/Fe] = $-$0.59. Cr abundances measured using Cr II lines 
whenever possible also show similar trends.

Mn abundance is calculated using spectrum synthesis calculation of 
6013.51 {\rm \AA} line considering  hyperfine structures  from 
Prochaska \& McWilliam (2000).
Except for HD~16458, that shows near solar abundance with [Mn/Fe] ${\sim}$ 0.06,
 the rest  show  underabundance  of Mn with [Mn/Fe] $\le$ -0.22. 

 Except for HD~16458 with [Co/Fe] ${\sim}$ 0.49, all other
stars in our sample show near-solar values for Co. 

 Abundances of Ni measured from Ni I lines give
near-solar values for all the stars. 

  HD~16458 and 5395 show mild overabundance with [Zn/Fe] ${\sim}$ 0.43 
and 0.29 respectively.  The rest of the objects show near solar values.

\subsection{ Sr, Y, Zr}
The Sr abundance was derived from Sr I line at 4607.327 {\rm \AA} 
for seven stars. HD~16458, 125079, 4395, 48565 and 216219 show 
overabundances  with  [Sr/Fe] $>$ 1.0.  The stars HD~5395 and 81192 show 
only a slight overabundances with [Sr/Fe] ${\sim}$ 0.26 and 0.58 respectively. 
For the remaining  four stars  the Sr abundance could not be measured as the  
line at  4607.327 {\rm \AA} appears distorted. None of 
the Sr II lines detected in the spectra were usable for
abundance estimates.

The abundance of Y was derived for all the stars except for  HD~201626.
HD~16458, 125079, 216219 and 48565 show an overabundance
 with [Y/Fe]$>$ 1.0. In the case of HD~4395, [Y/Fe] ${\sim}$ 0.65. 
The remaining  four stars show near-solar values. 

We could derive Zr abundance for six stars. Three stars show an overabundance 
with [Zr/Fe] $\ge$ 1.0.  HD~214714 shows an underabundance with [Zr/Fe] $\sim$  $-$0.28 and  HD~81192 gives a near-solar value. HD~4395 shows overabundance 
with  [Zr/Fe] ${\sim}$ 0.58.

\subsection{ Ba, La, Ce, Pr, Nd, Sm, Eu, Dy}

Standard abundance determination method using equivalent width measurements 
were used for  the elements Ce, Pr, Nd, Sm and Dy. Spectrum synthesis 
calculation was carried out for Ba, La and Eu.  We could determine the abundances 
for  Ba, Ce and Nd for all the stars. The derived abundances are found 
to be overabundant with respect to Fe. \\
Barium (Ba): We have determined {\bf Ba abundances} by synthesising  
Ba II line at 5853.668 ${\rm \AA} $ considering hyperfine 
components from Mcwilliam (1998). Five of our program stars HD 125079, 
216219, 16458, 48565 and 201626 show over abundance of 
Ba with [Ba/Fe] $\ge$ 1.0. HD 4395 shows an over abundance with 
[Ba/Fe] ${\sim}$ 0.79. While HD 5395, 81192 and 188650 show 
near solar values, HD 214714 shows mild under abundance with 
[Ba/Fe] ${\sim}$ $-$ 0.31.

Lanthanum (La): We derived La abundance for HD 16458, 
201626, 216219, 4395, 48565, 5395, 81192 from  spectrum 
synthesis calculation of La II line at 4921.77 ${\rm \AA}$ 
considering  hyperfine components from Jonsell et al. (2006). 
Except for HD~81192 and 5395, La in  all other stars is found to be  
overabundant relative to Fe with [La/Fe] $\ge$ 1. HD 81192 and 
5395 show  [La/Fe] of   $-$ 0.13 and 0.24 respectively.

Cerium (Ce): We derived Ce abundance for all the program stars. 
In HD~125079  [Ce/Fe] ${\sim}$ 0.93. While three stars HD~188650, 214714 
and 5395,  show almost near-solar  values, HD~16458, 201626, 216219 
and 48565  give 1.47, 1.89, 1.03 and 1.42 respectively for [Ce/Fe].
HD~4395  give  [Ce/Fe] ${\sim}$ 0.42 and  HD~81192 show a mild 
underabundance with [Ce/Fe] ${\sim}$ $-$0.15. 

Praseodymium (Pr):  We could derive Pr abundance in all the  program stars 
except for HD~81192 mainly using  the Pr II line at 5292.619 {\rm \AA}.
A mild over enhancement of Pr is seen in  HD~188650  and 4395 with 
[Pr/Fe] ${\sim}$ 0.57 and 0.53 respectively. In all other stars Pr shows 
an overabundance with values  ${\ge}$ 0.79. 

Neodymium (Nd): Abundance of  Nd was estimated for all the program stars. 
HD 188650 and 214714 show mild overabundance  with [Nd/Fe] ${\sim}$ 
0.39 and 0.36 respectively. Two stars HD~4395 and 5395 give 
[Nd/Fe] ${\sim}$ 0.8 and 0.74 respectively. While HD~16458, 48565, 
125079, 216219 and 81192 show an overabundance with 
[Nd/Fe] ${\ge}$ 1.0, HD~201626 shows a large overabundance with 
[Nd/Fe] ${\sim}$ 2.24.

Samarium (Sm):   Except for HD 125079, we used at least two clean lines 
for deriving the Sm abundance. HD~188650 shows a mild underabundance
with [Sm/Fe] ${\sim}$ $-$0.12. This value is found to be 0.56, 0.46, 0.85
and 0.91 respectively in HD~125079, 214714, 81192 and 216219.
Abundance of Sm is derived from a single Sm II line at 
4577.690 {\rm \AA} in HD~125079. Four objects HD~16458, 
201626, 4395 and 48565 show an overabundance with 
[Sm/Fe] ${\sim}$ 1.87, 1.63, 1.08 and 1.18 respectively.
We could not estimate Sm abundances in  HD~5395.

Europium (Eu): The abundance of Eu wass derived  for HD~16458 and 5395 
using  spectrum synthesis of Eu II lines at 6437.640 {\rm \AA} and 
6645.130 {\rm \AA} by considering  the hyperfine components from 
Worley et al. (2013). Eu shows an overabundance with 
[Eu/Fe] ${\sim}$ 0.67 and 0.34 respectively. In  HD 48565 we have used 
the Eu II line at 4129.700 {\rm \AA} and hyperfine components are 
taken from Mucciarelli et al. (2008). Eu is found to be slightly 
overabundant with [Eu/Fe] ${\sim}$ 0.29.  In case of HD~216219 we note that
the above  Eu II lines are  marginally asymmetric on the right wings.
These lines however, return  a near solar value  with [Eu/Fe] ${\sim}$  0.07
for this object.

Dysprosium (Dy): We could derive Dy abundance for three objects HD~201626, 
5395 and 81192 using the Dy II lines at  4103.310 {\rm \AA} and 
4923.167 {\rm \AA}. Dy shows an overabundance in these objects with 
[Dy/Fe] $\geq$ 1.0.
{\footnotesize
\begin{table*}\tiny
{\bf  Table 6: Elemental abundances}
\begin{tabular}{llllllllllllll}
\hline
Star Name&[Na I/Fe]&[Mg I/Fe]&[Ca I/Fe]&[Sc II/Fe]&[Ti I/Fe]&[Ti II/Fe]&[V I/Fe]&[Cr I/Fe]&[Cr II/Fe]&[Mn I/Fe]&[Co I/Fe]&[Ni I/Fe]&[Zn I/Fe] \\
\hline
Subgiant-CH stars\\
HD~4395     & 0.23  & 0.12 & 0.01 & -0.09& 0.04 & 0.22 & -0.14 & 0.01  & 0.15  &-0.23& -0.19& -0.02& 0.18\\
HD~125079   & 0.34  & 0.05 & 0.03 & 0.02&  0.0 & 0.45 & -0.12 & -0.06 & -    & -0.22 & 0.03 & 0.07 &  0.11\\
HD~216219   & 0.32  & 0.15 & 0.14 & -0.18   & 0.11 & 0.29 & -0.13  & -0.09 & 0.12  &-0.32 & 0.03 & 0.00 & 0.16\\
\hline
$\#$CH stars \\
$^*$HD~5395     & -0.39 & 0.12 & 0.04 &0.04& 0.08 & 0.18 & -0.24    & -0.19 &  0.0  &-0.5 & 0.20 & -0.09& 0.29 \\
$^*$HD~16458    & 0.65  & 0.04 & 0.28 & 0.10  & 0.31 & 0.41 & 0.25  & 0.29  & -    & 0.06 & 0.49 & 0.15 & 0.43\\
$^*$HD~48565    & 0.18  & 0.15 & 0.11 & -0.11 & 0.04 & 0.30& -0.21  & -0.19 & -0.32 & -0.42& -0.15& -0.17 & -0.18\\
HD~81192    &-0.29  & 0.29 & 0.18 &0.25    & 0.25 & 0.25 & 0.10  & -0.33 &  -0.23&-0.40& 0.28 & 0.08 & -0.14\\
HD~188650   & 0.76  & 0.27 & -0.02& -0.49& 0.24 & 0.24 & -0.24 & -0.04 & 0.01  &-0.24 & 0.06 & -0.16& -0.05\\ 
HD~214714   & 0.47  & 0.0  & 0.17 & -0.54& -0.19& 0.32 & -0.44  & -0.32 & -0.23 &-0.49& -0.06& -0.26&  -0.13\\
HD~201626   & 0.41  & 0.69 & 0.65 & 0.04   & 0.74 & 0.81 & -     & -0.59 & -    &- & 0.00 & -0.11& -0.04\\

\hline
\end{tabular}
$^\#$ Objects from the CH star catalogue of Bartkevicius (1996)\\
$^*$ Objects are also included in Ba star catalogue of L\"u (1991\\
\end{table*}
}
{\footnotesize
\begin{table*}\tiny
{\bf  Table 6: continued}\\
\begin{tabular}{llllllllllll}
\hline
Star Name& [Sr I/Fe]&[Y II/Fe]&[Zr II/Fe]&[Ba II/Fe]&[La II/Fe]&[Ce II/Fe]&[Pr II/Fe]&[Nd II/Fe]&[Sm II/Fe]&[Eu II/Fe]&[Dy II/Fe]\\
\hline
Subgiant-CH stars\\
HD 4395 &1.08& 0.65& 0.58& 0.79& 1.03&  0.42  &0.53  &0.80 & 1.08&  -&-\\
HD 125079& 1.59 &1.05& - &1.06&    - &0.93 & 1.00& 1.16& 0.56 &-&-\\
HD 216219 &1.8 & 1.00 &0.98 &1.10 & 1.04& 1.03& 1.14 &0.99 &0.91 & 0.07&-\\
\hline
$\#$CH stars\\
$^*$HD 5395 &0.26& 0.05& - &0.03 & 0.24 & 0.06& 0.79& 0.74 &-& 0.34&1.38\\
$^*$HD 16458 & 1.37 &1.46& 1.17& 1.18 &1.42& 1.47& 1.82 &1.55& 1.87& 0.66 &-\\
$^*$HD 48565& 1.73& 1.08 & 0.9 & 1.52 &1.46 &1.42 &1.29& 1.51 &1.18 & 0.29&-\\
HD 81192 &0.58& 0.10 &0.12 & 0.13 &  -0.13 &-0.15 &- &1.01 & 0.85 &-&1.21\\
HD 188650&- &-0.03 &-& -0.01 &- &-0.03& 0.57& 0.39 &-0.12& -&-\\
HD 214714 &-&0.22 & -0.28 &-0.31& -& 0.05& 0.93 &0.36& 0.46 &-&-\\
HD 201626& - &- &- &2.12& 1.76& 1.89& 2.09 &2.24& 1.63& -&0.97\\
\hline
\end{tabular}
$^\#$ Objects from the CH star catalogue of Bartkevicius (1996)\\
$^*$ Objects are also included in Ba star catalogue of L\"u (1991\\
\end{table*}
}
{\footnotesize
\begin{table*}
{Table 7: Observed values for [Fe/H], [ls/Fe], [hs/Fe]  and [hs/ls]}\\
\begin{tabular}{lcccccc}
\hline
Star Name & [Fe/H]  & [ls/Fe] & [hs/Fe] & [hs/ls] & Remarks   \\
\hline
HD~4395  & $-$0.18  & 0.77    &  0.82   &  0.05   &    1      \\
         & $-$0.33  & 0.7     &  0.5    & $-$0.2  &    2       \\
         &          &         &         &         &           \\
HD~5395  & $-$0.24  & 0.16    & 0.27    & 0.11    &    1     \\
         &          &         &         &         &           \\
HD~16458 &  $-$0.65   & 1.34    &  1.50   &  0.16   &  1        \\
         &          &         &         &         &           \\
HD~48565 & $-$0.59  & 1.24    & 1.47    & 0.23     &     1     \\
         &          &         &         &         &           \\
HD~81192 & $-$0.50  & 0.26    & 0.34    & 0.08    &    1    \\
         &          &         &         &         &           \\
HD~125079&  $-$0.18 &  1.32   & 0.93   &  $-$0.39  &   1     \\
         &          &         &         &         &           \\
HD~188650 & $-$0.45  & $-$0.03& 0.23    & 0.26    &   1      \\
         &          &         &         &         &           \\
HD~201626& $-$1.39  &  -      & 1.93    &  -      &   1      \\
         & $-$1.3   & 1.1     &  1.6    & 0.5     &   2       \\
         &          &         &         &         &           \\
HD~214714& $-$0.35  & $-$0.03 & 0.14    &  0.17   &   1      \\
         &          &         &         &         &           \\
HD~216219& $-$0.17  & 1.26    & 1.01    & $-$0.25 &    1      \\
         & $-$0.32  & 1.0     &  0.9    & $-$0.1  &    2       \\
\hline
\end{tabular}

1. Our work; 2: Busso et al. (2001) \\
\end{table*}
}

\begin{table*}\tiny
{\bf  Table 8: Equivalent widths (in m{\rm \AA},) of lines used for the calculation of light element abundances }\\
\begin{tabular}{lllcllllllllll}
\hline
Wavelength& Element &    E$_{low}$ &   log gf& HD~4395&HD~5395 & HD~16458&HD~48565&HD~81192& HD~125079&HD~188650&HD~201626&HD~214714&HD~216219\\
\hline
 5682.650 &  Na I  & 2.100  & -0.700   &  94.6     &     - &   143.4          &   48.66    &     -&  109.7    &   109.9   &   -      &  95.46  &  81.17 \\
 5688.220 &        & 2.100  & -0.400   &  100.6     &    -     &    149.9   &  -         &     -&  120.9    &   138.2&   -      &  128.4  &	 107.1 \\
 5889.950 &        & 0.000  &  0.100   &  -       &    411.5 &   -       &  -     &     411.5   &  -        &   -    &   -      &  -      &	 358.0 \\
 5895.920 &        & 0.000  & -0.200   &  347.2      &    378.6 &   536.7   &  242.5     &     378.6   &  -        &   -    &   214.0  &  -      &	 284.4 \\
 4702.990 &  Mg I  & 4.350  & -0.666   &  -          &    -     &   238.8   &  150.3     &     202.1   &  199.9    &   174.9&   133.2  &  174.9  &	 178.8 \\
 6318.720 &        & 5.108  & -1.730   &  -          &    52.21 &   52.79   &  15.55     &     -&  -        &   -    &   -      &  26.95  &	 -     \\
 5528.400 &        & 4.350  & -0.490   &  185.7      &    211.8 &   192.4   &  143.5     &     211.8   &  199.5    &   184.9&   -      &  176.2  &	 157.2 \\
 4098.520 &  Ca I  & 2.525  & -0.540   &  -          &    -     &   -       &  47.8     &     -&  -        &   66.2&   63.5  &  95.4  &	 85.3 \\
 4283.010 &        & 1.885  & -0.224   &  -          &    -     &   -       &  109.8     &     -&  -        &   -    &   -      &  -      &	 149.9 \\
 4425.430 &        & 1.879  & -0.385   &  136.8      &    -     &   102.2   &  99.6     &     144.0   &  -        &   132.1&   -      &  -      &	 114.5 \\
 4435.679 &        & 1.890  & -0.520   &  133.6      &    -     &   -       &  109.5     &     141.0   &  -        &   -    &   -      &  137.3  &	 122.1 \\
 4455.890 &        & 1.900  & -0.510   &  137.6      &    -     &   -       &  104.3     &     137.7   &  165.7    &   132.9&   107.8  &  113.2  &	 133.7 \\
 4578.550 &        & 2.521  & -0.560   &  64.2      &    -     &   -       &  47.19     &     102.6   &  98.7    &   47.52&   -      &  -      &	 67.2 \\
 5261.710 &        & 2.521  & -0.730   &  90.9      &    112.4 &   146.6   &  55.70     &     -&  102.7    &   80.50&   52.2  &  92.9  &	 79.5 \\
 5265.560 &        & 2.520  & -0.110   &   -         &    -&   -  &  -         &      -      & 	-    &    -      &    160.  & 5 -     &	  -    \\
 5512.990 &        & 2.932  & -0.290   &  76.5      &    -     &   135.3   &  45.36     &     87.4   &  86.2    &   54.8&   -      &  -      &	 68.1 \\
 5581.980 &        & 2.523  &  -0.710  &  89.1      &    -     &   125.7   &  58.81     &     115.6   &  107.2    &   79.45&   -      &  93.9  &	 84.51 \\
 5588.760 &        & 2.530  &  0.360   &  129.7      &    -     &   -       &  102.8     &     140.9   &  141.8    &   137.4&   -      &  137.5  &	 127.7 \\
 5590.130 &        & 2.521  & -0.710   &  82.7      &    -     &   94.4   &  54.1     &     103.2   &  94.0    &   -    &   -      &  86.08  &	 77.9 \\
 5594.470 &        & 2.520  & 0.100    &  108.9      &          &   -     &  -        &-&  156.0    &   -    &   164.6  &  -      &	 -     \\
 5857.450 &        & 2.930  & 0.240    &  120.8      &    -     &   170.3   &  82.3     &     131.4   &  137.5    &   130.9&   -      &  -      &	 111.0 \\
 6162.170 &        & 1.900  & -0.090   &  -          &    210.5 &   -       &  -         &     200.9   &  204.8    &   187.3&   132.0  &  185.2  &	 162.9 \\
 6439.070 &        & 2.530  & 0.390    &  141.8      &    168.0 &   185.2   &  106.8     &     161.3   &  162.8    &   162.6&   -      &  161.8  &	 140.0 \\
 6449.820 &        & 2.520  & -0.550   &  78.5      &    -     &   -       &  60.7     &     -&  -        &   -    &   -      &  118.8  &	 88.7 \\
 6455.590 &        & 2.523  & -1.350   &  53.9      &    -     &   111.0   &  -         &    79.7&  -        &   -    &   -      &  -      &	 45.5 \\
 6471.670 &        & 2.525  & -0.590   &  93.7      &    -     &   134.8   &  52.3     &     -&  98.8   &   81.1&   54.5  &  86.4  &	 79.6 \\
 6493.790 &        & 2.521  &  0.140   &  145.8      &    -     &   168.4   &  88.1     &     136.8   &  130.6    &   111.4&   -      &  127.0  &	 111.9 \\
 6499.650 &        & 2.523  & -0.590   &  79.7      &    -     &   -       &  44.3     &     -&  95.9    &   55.5&   -      &  -      &	 75.1 \\
 6717.690 &        & 2.709  & -0.610   &  -          &    -     &   186.0   &  60.7     &     -&  -        &   -    &   -      &  105.6  &	 95.9 \\
 4431.110 &        & 6.035  & -2.520   &  -          &    -     &   25.0   &  -         &     -&  -        &   -    &   68.3  &  -      &	 -     \\
 4400.389 &  Sc II & 0.606  & -0.510   &  -          &    –&   -  &  -         &    &  -        &   -    &   -      &  158.5  &	 -     \\
 5031.021 &        & 1.360  & -0.260   &  79.5      &    -     &   -       &  61.9     &     -&  119.3    &   131.7&   -      &  -      &	 -     \\
 6604.600 &        & 1.357  & -1.480   &  -          &    -     &   -       &  22.7     &     -&  60.41    &   -    &   -      &  65.85  &	 -     \\
 6245.637 &        &1.507   & -1.030   & -& 73.1      & 74.5      & 21.7      & 61.4       &56.9       &69.3       &34.3       &72.7      & 46.9\\
 4512.700 & Ti I   & 0.836  & -0.480   &  64.9      &    -     &   115.3   &  34.9     &     79.1   &  82.9    &   48.3&   -      &  56.7  &	 53.7 \\
 4453.710 &        & 1.873  & -0.010   &  52.9      &    -     &   -       &  18.1    &     66.57   &  -        &   -    &   -      &  -      &	 -     \\
 4533.239 &        & 0.848  &  0.476   &  111.6      &    -     &   -       &  71.4     &     140.1   &  -        &   -    &   73.6 &  -      &	 99.25 \\
 4617.250 &        & 1.748  &  0.389   &  -          &    88.4 &   -       &  29.3     &     83.5   &  -        &   35.31&   -      &  -      &	 -     \\
 4656.468 &        & 0.000  & -1.345   &  56.2     &    -     &   150.1   &  30.4     &     104.7   &  85.8    &   37.8&   -      &  46.8  &	 -     \\
 4759.272 &        & 2.255  &  0.514   &  -          &    76.7 &   -       &  18.3     &     63.9   &  59.7    &   16.4&   -      &  26.4  &	 39.2 \\
 4820.410 &        & 1.502  & -0.441   &  -          &    89.2 &   -       &  -         &     86.5   &  -        &   23.6&   64.5  &  36.9  &	 62.7 \\
 4840.880 &        & 0.899  & -0.509   &  -          &    -     &   119.3   &  26.9     &     86.4   &  79.4    &   -    &   -      &  -      &	 52.7 \\
 4778.250 &        & 2.236  & -0.220   &  -          &    34.3 &   71.7   &  -         &     -&  19.6    &   -    &   -      &  -      &	 -     \\
 4999.500 &        & 0.830  & 0.310    &  101.0      &    -     &   -       &  -         &     136.7   &  -        &   -    &   -      &  131.2  &	 -     \\
 4937.730 &        & 0.813  & -2.230   &  117.1      &    35.1 &   74.1   &  -         &     -&  -        &   -    &   -      &  -      &	 -     \\
 5007.210 &        & 0.820  & 0.170    &  -          &    -     &   -       &  -         &     152.3   &  -        &   -    &   108.8  &  -      &	 -     \\
 5039.960 &        & 0.020  & -1.130   &  -          &    124.6 &   -       &  42.8     &     -&  105.3    &   66.9&   -      &  83.4 &	 72.9 \\
 5064.650 &        & 0.050  & -0.940   &  94.0      &    145.3 &   -       &  42.8     &     -&  -        &   -    &   -      &  -      &	 68.9 \\
 5087.060 &        & 1.429  & -0.780   &  -          &    80.2 &   93.8   &  -         &     92.8   &  -        &   -    &   -      &  -      &	 -     \\
 5210.390 &        & 0.048  & -0.884   &  -          &    127.5 &   -       &  52.7     &     -&  111.4    &   81.2&   -      &  97.3  &	 78.2 \\
 6556.060 &        & 1.460  & -1.074   &  93.1      &    60.7 &   104.0   &  -         &     -&  33.72   &   -    &   -      &  -      &	 -     \\
 4161.530 &  Ti II & 1.080  & -2.160   &  66.2      &    -&   - -&  -         &    &  -        &   -    &   -      &  108.1  &	 -     \\
 4417.710 &         & 1.160 & -1.430   &  -          &    -     &   -       &  -         &     -&  141.8    &   -    &   -      &  160.8  &	 117.5 \\
 4418.330 &         & 1.240 & -1.990   &  -          &    -     &   140.0   &  67.5    &    98.3   &  -     &   -    &   -      &  112.7  &	 83.42 \\
 4443.790 &         & 1.080 &  -0.700  &  138.1      &    175.3 &   179.7  &  123.4     &     124.6   &  181.5    & 196.8&   128.6  &  -      &	 145.3 \\
 4468.520 &         & 1.130 & -0.600   &  -          &    -     &   -       &  -         &     -&  186.8    &   -    &   -      &  -      &	 -     \\
 4470.900 &         & 1.164 & -2.280   &  -          &    88.9 &   -       &  53.8     &     -&  -        &   115.7&   71.01  &  -      &	 75.79 \\
 4493.510 &         & 1.080 &  -2.730  &  -          &    -   &   84.4   &  32.04     &     52.10   &  -     &   62.26&   60.59  &  79.18  &	 60.06 \\
 4563.760 &         & 1.221 & -0.960   &  131.2      &    -  &   210.7   &  110.4     &     150.6   &  -      &   -    &   121.9  &  193.6  &	 141.6 \\
 4571.960 &         & 1.571 & -0.530   &  139.1      &    -     &   -       &  118.5     &     -&  -     &   -    &   68.33  &  -      &	 165.1 \\
 4568.310 &         & 1.224 &  -2.650  &  -          &    -     &   94.1   &  -         &     66.1   &  -        &   56.8&   15.3  &  -      &	 35.8 \\
 4657.210 &         & 1.240 &  -2.320  &  51.7     &    -     &   47.41   &  46.81     &     64.6   &  -        &   -    &   -      &  -      &	 -     \\
 4708.665 &         & 1.240 &  -2.370  &  -          &    83.6 &   -       &  -         &     -&  -        &   -    &   -      &  -      &	 66.7 \\
 4798.510 &         & 1.080 & -2.670   &  -          &    -     &   -       &  32.8     &     -&  -        &   -    &   -      &  -      &	 65.8 \\
 4805.090 &         & 2.060 & -1.100   &  101.1      &    -   &   134.3   &  75.2     &     108.5   &  -        &   154.6&   100.8  &  147.1  &	 108.2 \\
 5185.900 &         & 1.890 &  -1.350  &  -          &    -&   - -&  88.3     &    &  -        &   63.9&   126.3  &  -      &	 69.8 \\
 5226.530 &         & 1.570 & -1.300   &  -          &    141.9 &   165.2   &  83.7     &     107.8   &  -        &   -    &   96.5 &  -      &	 110.1 \\
 4090.570 &  V I    & 1.853 &-2.099    &  -          &    -&   - -&  -         &    &  -        &   -    &   96.4  &  -      &	 -     \\
 4379.230 &         & 0.300 &  0.580   &  119.3      &    - &   215.3   &  57.40     &     110.8   &  124.3    &   84.8&   -      &  -      &	 87.3 \\
 4406.630 &         & 0.300 & -0.190   &  -          &    -     &   220.7   &  47.2    &     122.9   &  -        &   73.4&   -      &  -      &	 68.1 \\
 4876.430 &         & 2.115 & -2.714   &  -          &    -     &   -       &  -         &     -&  -        &   -    &   -      &  99.8  &	 -     \\
 5727.653 &         & 1.051 &  -0.870  & -           & 95.0     & 76.3     &45.0      & 37.1    & 17.2       & -     &-         & -      & -           \\
 6531.420 &         & 1.218 & -0.840   &  -          &    -     &   51.1  &  -         &     -&  -        &   -    &   -      &  -      &	 -     \\
 4274.800 &   Cr I     & 0.000 & -0.230   &  200.8      &    -     &   -       &  147.4     &     -&  -        &   -    &   -      &  -      &	 161.6 \\
 4289.720 &         & 0.000 &-0.360    &  -          &    -     &   -       &  86.5     &     -&  -        &   -    &   88.4  &  -      &	 -     \\
 4351.050 &         & 0.970 &  -1.450  &  55.1     &    -     &   -       &  -         &     48.6   &  -        &   -    &   -      &  -      &	 45.31 \\
 4600.750 &         & 1.000 & -1.260   &  -          &    -     &   -       &  -         &     -&  -        &   -    &   -      &  -      &	 47.8\\
 4616.140 &         & 0.980 & -1.190   &  87.4      &    115.9 &   202.5   &  54.7     &     -&  110.1    &   99.9&   32.8  &  116.4  &	 74.0 \\
 4626.190 &         & 0.970 & -1.320   &  80.9      &    106.9 &   111.6   &  43.8     &     94.8   &  97.7    &   86.3&   -      &  100.9  &	 62.25 \\
 4652.160 &         & 1.000 & -1.030   &  103.1      &    -     &   184.1   &  57.6     &     113.2   &  122.2    &   110.8&   -      &  117.4  &	 94.59 \\
 4737.380 &         & 3.087 & -0.099   &  59.9      &    75.8 &   109.8   &  -         &     -&  -        &   24.3&   -      &  -      &	 68.6 \\
 4942.490 &        & 0.941  & -2.294   &  -          &    -     &   163.5   &  29.2     &     -&  -        &   64.8&   -      &  72.1  &	 -     \\
 5206.040 &        & 0.940  &  0.020   &  -          &    -     &   -       &  -         &     -&  -        &   -    &   -      &  -      &	 169.3 \\
 5247.570 &        & 0.961  & -1.640   &  77.4      &    -     &   140.5   &  34.3     &     84.7   &  -        &   77.2&   -      &  89.1  &	 641 \\
 5345.810 &        & 1.003  &  -0.980  &  115.4      &    159.3 &   -       &  70.7     &     140.8   &  137.9    &   -    &   -      &  140.2  &	 105.8 \\
 5348.312 &        & 1.003  &  -1.290  &  -          &    -     &   -       &  -         &     -&  -        &   -    &   51.3  &  107.0  &	 -     \\
\hline
\end{tabular}
\end{table*}

 \begin{table*}\tiny
{\bf  Table 8: Continued }\\
\begin{tabular}{lllcllllllllll}
\hline
Wavelength& Element &    E$_{low}$ &   log gf& HD~4395&HD~5395 & HD~16458&HD~48565&HD~81192& HD~125079&HD~188650&HD~201626&HD~214714&HD~216219\\

\hline
 
 4588.190 &  Cr II & 4.072  & -0.630   &  66.6      &    73.5 &   -       & 57.7      &     49.3   &  –        &   131.1&   -      &  99.9  &	 84.2 \\
 4592.040 &        & 4.073  & -1.220   &  33.2      &    48.6 &   -       & 33.2      &     27.3   &  –        &   81.9 &   -      &  64.2  &	 52.7 \\
 4812.350 &        & 3.864  & -1.800   &  -          &    -     &   -       & 19.8     &     40.9   &  –        &   -    &   -      &  42.5  &	 41.67 \\
 4634.070 &        & 4.073  & -1.240   &  -          &    -     &   -       & -          &     -&  -        &   102.5&   -      &  -      &	 73.7 \\
 4848.250 &        & 3.864  & -1.140   &  -          &    -     &   -       & -          &     -&  -        &   -    &   -      &  87.16  &	 65.3 \\
 4030.750 &  Mn I& 0.000    & -0.470   &  -          &    -     &   -       & -          &     -&  -        &   -    &   115.1  &  -      &	 - \\
 4034.480 &        & 0.000  & -0.810   &  -          &    -     &   -       & 125.3      &     -&  -        &   176.0&   -      &  -      &	 183.0 \\
 4041.360 &        & 2.110  &  0.280   &  -          &    195.2 &   -       & 102.6      &     -&  151.3    &   -    &   68.66  &  -      &	 143.0 \\
 4451.580 &        & 2.890  &  0.280   &  -          &    -     &   131.7   & 73.42      &     73.69   &  116.5    &   95.92&   -      &  98.18  &	 97.47 \\
 4470.140 &        & 2.941  & -0.444   &  37.5      &    56.8 &   59.4   & 11.5      &     41.4   &  63.6    &   41.2&   23.1  &  23.8  &	 28.00 \\
 4739.080 &        & 2.941  & -0.490   &  44.2      &    80.9 &   124.6   & 17.9      &     -&  73.4    &   48.3&   -      &  56.2  &	 36.9 \\
 4761.530 &        & 2.953  &  -0.138  &  67.8      &    100.0 &   105.9   & 25.3      &     -&  90.2    &   62.8&   -      &  68.3  &	 59.6 \\
 4765.860 &        & 2.941  & -0.080   &  68.4      &    100.8 &   95.9   & 32.8      &     -&  82.8    &   63.5&   -      &  67.0 &	 57.0 \\
 4766.420 &        & 2.919  &   0.100  &  83.0      &    127.8 &   140.7   & 46.38      &     109.6   &  -        &   95.96&   23.9  &  -      &	 72.52 \\
 4783.430 &        & 2.300  &  0.040   &  125.6      &    159.3 &   200.9   & 75.8      &     -&  137.9    &   136.8&   -      &  138.1  &	 107.2 \\
 5516.770 &        & 2.178  &  -1.847  &  -          &    93.2 &   141.6   & -          &     -&  -        &   26.2&   -      &  43.52  &	 20.2 \\
 6013.488 &        &  3.072 & -0.250   &75.8         &  108.3   &   148.9   &   23.9   &   84.9   &   98.9   &   66.9    &  -   &  74.8     &  54.3  \\
 4813.480 &  Co I & 3.216   &   0.050  &  -          &    -     &   -       & 10.57      &     46.3   &  -        &   24.2&   -      &  -      &	 -     \\
 4121.320 &        & 0.920  &  -0.320  &  125.2      &    -     &   -       & -          &     -&  -        &   -    &   -      &  135.3  &	 107.9 \\
 4118.770 &        & 1.050  &  -0.490  &   -         &     –    &    -      & -          &&  -        &   -    &   70.1  &  -      &	 -     \\
 4749.660 &        & 3.053  &  -0.321  &  -          &    -     &   -       & -          &     74.8   &  -        &   -    &   -      &  -      &	 -     \\
 4771.080 &        & 3.133  &  -0.504  &   -         &     59.0 &    72.4  & -          &     -&  -        &   -    &   -      &  -      &	 -     \\
 4781.430 &        & 1.882  &  -2.150  &  -          &    33.8 &   73.7   & -          &     -&  -        &   -    &   -      &  -      &	 -     \\
 4792.850 &        & 3.252  &  -0.067  &  29.3     &    52.6 &   -       & -          &     -&  -        &   -    &   -      &  -      &	 -     \\
 5530.770 &        & 1.710  &  -2.060  &  -          &    52.9 &   98.5   & -          &     46.1   &  -        &   -    &   -      &  -      &	 -     \\
 5590.720 &        & 2.042  &  -1.870  &  -          &    47.3 &   92.1   & -          &     -&  25.2    &   -    &   -      &  -      &	 -     \\
 5991.870 &        & 2.080  &  -1.850  &  -          &    59.7 &   97.7   & 7.483      &     -&  -        &   -    &   -      &  -      &	 -     \\
 6454.990 &        & 3.632  &  -0.233  &  -          &    -     &   40.2   & -          &     -&  -        &   -    &   -      &  -      &	 -     \\
 6632.430 &        & 2.280  &  -2.000  &  -          &    29.3 &   59.6   & -          &     -&  -        &   -    &   -      &  -      &	 -     \\
 4470.470 &  Ni I  & 3.699  & -0.400   &  -          &    -     &   -       & -          &     -&  -        &   -    &   -      &  34.3  &	 -     \\
 4714.420 &        & 3.380  &  0.230   &  -          &    -     &   -       & 83.4      &     132.7   &  164.4    &   -    &   -      &  -      &	 126.4 \\
 4732.460 &        & 4.106  & -0.550   &  33.18      &    -     &   -       & 11.5      &     29.5   &  -        &   -    &   -      &  -      &	 -     \\
 4814.590 &        & 3.597  & -1.680   &  -          &    -     &   54.4   & -          &     16.3   &  49.7    &   -    &   -      &  -      &	 -     \\
 4821.130 &        & 4.153  & -0.850   &  -          &    -     &   60.1   & -          &     32.8   &  -        &   15.4&   -      &  -      &	 -     \\
 4852.560 &        & 3.542  & -1.070   &  32.5      &    -     &   -       & -          &     46.8   &  52.1    &   24.4&   -      &  32.7  &	 28.3 \\
 4855.410 &        & 3.540  &  0.000   &  84.8      &    -     &   111.1   & 53.2      &     98.9   &  -        &   -    &   27.5  &  -      &	 -     \\
 4857.390 &        & 3.740  & -1.199   &  44.6      &    -     &   -       & 13.5      &     60.7   &  60.2    &   27.8&   -      &  -      &	 34.2 \\
 4953.200 &        & 3.740  & -0.670   &  46.9      &    -     &   -       & 17.8      &     66.7   &  61.6    &   36.9&   -      &  38.8  &	 35.6 \\
 4937.340 &        & 3.606  &  -0.390  &  66.6      &    -     &   120.2   & 36.3      &     -&  -        &   62.7&   -      &  71.8  &	 61.9\\
 4980.160 &        & 3.610  &  -0.110  &  94.5      &    -     &   -       & 57.5      &     92.2   &  -        &   -    &   40.1  &  92.6  &	 87.8 \\
 5035.370 &        & 3.630  &   0.290  &  79.3      &    -     &   94.3   & 63.6      &     94.6   &  -        &   95.4&   -      &  87.5  &	 73.7 \\
 5081.120 &        & 3.847  &  0.300   &  84.3      &    102.9 &   97.3   & 58.2      &     81.5   &  92.3    &   83.9&   -      &  80.6  &	 69.57 \\
 5082.350 &        & 3.657  & -0.540   &  62.5      &    76.6 &   -       & 30.0      &     -&  75.1    &   42.3&   -      &  58.9  &	 47.3 \\
 5084.080 &        & 3.678  &  0.030   &  74.9      &    94.1 &   77.7  & 51.8      &     75.9   &  90.2    &   72.5&   -      &  74.1  &	 71.5 \\
 5099.930 &        & 3.678  & -0.100   &  80.9      &    -     &   -       & 46.8      &     -&  -        &   -    &   -      &  96.5  &	 73.6\\
 5102.960 &        & 1.676  & -2.620   &  58.7      &    -     &   -       & 18.2      &     -&  -        &   -    &   -      &  37.0  &	 35.4 \\
 5259.470 &        & 3.740  & -1.502   &  10.0      &    -     &   114.3   & -          &     -&  -        &   -    &   -      &  -      &	 17.8 \\
 6086.280 &        & 4.266  & -0.530   &  35.1      &    51.9 &   64.3   & 11.7      &     33.6   &  45.8    &   23.9&   -      &  27.2  &	 -     \\
 6111.070 &        & 4.088  & -0.785   &  31.6      &    45.5 &   59.1   & -     &     30.0   &  36.3    &   -    &   -      &  -      &	 27.4 \\
 6176.810 &        & 4.088  & -0.148   &  -          &    -     &   93.8   & 23.5      &     -&  68.6    &   42.0&   -      &  -      &	 20.3 \\
 6186.710 &        & 4.106  & -0.777   &  -          &    42.2 &   -       & -          &     -&  33.6    &   -    &   -      &  -      &	 48.2 \\
 6204.600 &        & 4.088  & -1.060   &  -          &    30.7 &   56.9   & -          &     -&  26.7    &   -    &   -      &  -      &	 -     \\
 6643.640 &        & 1.676  & -2.300   &  94.4      &    145.5 &   169.0   & 41.8      &     118.4   &  -        &   86.8&   50.9  &  100.9  &	 77.7 \\
 4722.150 &Zn I   & 4.029  &  -0.370  &  76.0      &    -     &   -       & 52.6      &     51.6   &  -        &   89.5&   152.3  &  91.2  &	 85.9 \\
 4810.530 &       & 4.080   &  -0.170  &  -          &    -     &   90.7   & 51.3      &     68.1   &  87.5    &   101.2&   47.1  &  89.5  &	 76.7 \\
\hline
\end{tabular}
\end{table*}

\begin{table*}\tiny
{\bf  Table 9: Equivalent widths (in m{\rm \AA},) of lines used for the calculation of heavy element abundances  }\\
\begin{tabular}{lllcllllllllll}
\hline
Wavelength& Element &    E$_{low}$ &   log gf& HD~4395&HD~5395 & HD~16458&HD~48565&HD~81192& HD~125079&HD~188650&HD~201626&HD~214714&HD~216219\\

\hline
 
 4607.327  &  Sr I&     0.000  &   -0.570  &60.4  &65.6  &  138.6  & 56.5 & 62.9& 95.9&    -    &      -      &  -        & 80.2 \\     
 4854.863  & Y II &    0.992  &   -0.380  &       &77.1  & 	  &   -   & 63.7&    -     &  95.3   &	-    &	-        & 98.1  \\  
 4883.685  &       &    1.084  &    0.07   &90.4  &81.1  &  211.9  & 87.2 & 72.4& 134.4 &  126.9 &	-    & 122.8     & 119.9 \\ 
 5087.416  &       &    1.080  &   -0.170  & 71.9 &71.3  & 	  & 80.2 & 71.7& 104.1    &  96.2   &	-    &	         & 98.1 \\ 
 5119.112  &       &    0.992  &  -1.360   &       &31.7  &  120.6  & 36.7 &   -  & 71.7    &  44.3   &	-    &	-        & 64.8 \\ 
 5205.724  &       &    1.033  &  -0.340   &       &   -   & 	  &   -   &  -   &    -     &    -     &	-    &	-        & 116.2 \\ 
 5289.815  &       &    1.033  &   -1.850  & 15.5 &   -   &  87.5  & 13.7 &  -   & 41.4    &    -     &	 -   &	-        & 34.2 \\ 
 5544.611  &       &    1.738  &  -1.090   &       &   -   & 	  & 22.09 &  -   &    -     &    -     &	-    &	-        & 41.6  \\
 5546.009  &       &    1.748  &  -1.100   & 30.9 &   -   &   27.9 &  -   &    -     &  22.0   &	 -       & -&22.2    & 51.6  \\
 5662.925  &       &    1.944  &    0.160  &       &   -   & 	  & 63.85 &  -   &    -     &    -     &	-    &	-	 &      - \\
 6613.733  &       &    1.748  &   -1.11   &       &42.6  &  119.5  & 22.3 &   -  &    -     &    -     &	-    &	-	 & 45.5  \\
 4048.670  &  Zr II &    0.800  &  -0.480   &       &   -   &  	  &   -   &  -   & -    &    -     &	-    &	-	 &      - \\
 4208.990  &        &    0.710  &  -0.460   & 68.2 &   -   & 	  & 68.2 & 64.9&    -     &    -     &	-    & 	-	 & 88.0  \\
 4317.321  &        &    0.713  &  -1.380   &       &   -   & 	  &     - &  -   &    -     &    -     &	-    & 30.3     & 66.0  \\
 4554.036  &  Ba II &    0.000  &    0.120  & 259.1 & 215.0 &- 	  &    -  & 176.2 & 412.3    &  284.4   &286.0    & 259.5     & 338.7    \\
 4130.650  &        &    2.720  &   +0.560  & 77.8 &   -   &  182.2  &   -   &  -   & 98.96    &    -     &	-    &	-	 & 108.8  \\
 4934.076  &        &    0.000  &  -0.150   &       &   -   & 	  & 265.6 &  -   &    -     &  298.1   &333.9    &	-	 &      - \\
 5853.668   &       &    0.604  &  -1.020   &       &   -   &  263.9  &   -   &  -   &    -     &  147.8   &	-    &	-	 &      - \\
 6141.727  &        &    0.704  &  -0.076   &       & 158.9 &  420.5  & 200.8 &   -  & 253.8    &  237.9   &	-    &	-	 & 230.0  \\
 6496.897  &         &    0.604  &  -0.377   &157.1  & 149.8 &  402.3  & 214.7 &   -  & 233.9    &  225.0   & 	-    & 229.9     & 209.4  \\
 4123.230  & La II &    0.320  &  +0.120   &       &   -   &  	  & 87.1 &  -   &    -     &    -     &	-    &	-	 &      - \\
 4238.380  &        &    0.400  & -0.058    & 92.5 & 68.0 & 	  & 101.1 &   -  &    -     &    -     &	-    &	-	 & 97.97  \\
 4322.510  &       &    0.170  &  -1.050   & 44.3 &  -    &         &  -    & 19.9&    -     &    -     &      -  &  -        &      -   \\
 4333.760  &        &    0.170  &  -0.160   & 108.8 &  -    &         & -     &56.6 &    -     &    -     &      -  &  -        &      -   \\
 4619.874  &      &    1.754  &  -0.140   &   -    &   -   & 	  & 22.8 &  -   &    -     &    -     &	-    &	-	 & 25.7  \\
 4748.726  &      &    0.927  &  -0.860   &    -   &   -   &  72.1 & 23.2 &  -   &    -     &    -     &	 -   & -    & 42.1  \\
 4921.776  &      &    0.244  &  -0.680   &  67.4  &   85.1 & 173.9& 64.5 & 72.9 & -        &    -     &   78.8      &-        & 77.9\\
 5808.313  &      &    0.000  &  -2.200   &    -   &   -   &  77.3  &   -   &  -   &    -     &    -     & 	-    &	-	 &      - \\
 6320.376  &      &   0.170   & -1.610   &     -  &   -   &         &   -   &  -   &          &   -      &54.4    &	-	 &      - \\
 4073.470  &  Ce II &    0.480  &  +0.320   &  -     &   -   & 	  & 60.5 &  -   & 58.5    &    -     &	     &	-	 & 61.5  \\
 4117.290  &        &    0.740  &  -0.450   &    -   &  -    &  39.2  &   -   &      &    -     &    -     &36.8   &-	         & 27.9  \\   
 4190.630  &         &    0.900  &  -0.390   &  -     &   -   & 	  & 19.1 &  -   &    -     &    -     &	 -   &	-	 &      -   \\
 4193.870  &         &    0.550  &  -0.400   &    -   &   -   & 	  & 33.5 &  -   &    -     &    -     &	 -   &	-	 &      - \\    
 4257.120  &         &    0.460  &  -1.116   &    -   &   -   &         &  -    &	 &    -     &    -     &  31.8  & -	 &      -    \\
 4336.244  &         &    0.704  & -0.564    &    -   &   -   & 	  &   -   &  -   &    -     &    -     &	 -   &	-	 & 34.9 \\  
 4418.790  &         &    0.860  &  +0.310   &    -   &   -   & 	  &   -   &  -   &    -     &  28.8   & 	 -   &	30.6    & 55.3 \\  
 4427.920  &         &    0.540  &  -0.380   &    -   &   -   & 	  &   -   &  -   &    -     &    -     &	 -   &   22.1   & 39.7  \\        
 4460.207  &          &    0.477  &   0.171   &   -    &   -   & 	  &   -   &  -   &    -     &  94.8   &	 -   &	-	 &      - \\
 4560.280  &          &    0.910  &   0.000   &   -    &   -   & 	  &  52.2 &  -   &    -     &    -     &	     &	-	 &      - \\
 4562.359  &         &    0.478  &   0.081   &    -   &   -   &  132.0  &  60.6 &34.1 & 74.9    &  48.6   &78.6    &  69.8    & 73.3    \\
 4628.160  &        &    0.520  &  +0.260   &     -  & 46.3 &  135.4  &  64.5 &   -  &    -     &  44.4   &	     &	-	 & 76.08  \\
 4747.167  &        &    0.320  &  -1.246   &     -  &   -   & 	  &  14.3 &  -   &    -     &    -     &	     &	-	 &      - \\
 4773.941  &        &    0.924  & -0.498    &     -  & 24.7 & 	   20.5 &   -  &   34.9    &    -     &47.4    & 12.3     & 30.5   &  \\
 4873.999   &       &   1.107  &  -0.892   &     -  &   -   &  57.4  &   -   &  -   &    -     &    -     & 	  -  &		 &      -  \\
 5274.230  &         &    1.044  &   0.323   &25.3  & 23.6 &  	  &  37.0 &   -  &    -     &    -     &	 -   & 23.9     & 45.2\\  
 5330.556  &         &    0.869  &   -0.760  &12.7  &   -   & 	  & 22.4 &  -   &     -     &    -     & 48.9   & 12.4     & 28.2     \\
 5188.217   & Pr II&     0.922  &  -1.145   &       &   -   &  34.1 &   -   &  -   &    -     &    -     &	  -  &	-	 &      - \\
 5259.740  &                 &   0.630  &  -0.070    & 10.4 &   -   &  104.6  &  15.1 &  -   &    -     &    -     &	  -  &	-	 &      - \\
 5219.045  &                 &    0.795  &  -0.240   &  -     &   -   &  92.5  &   -   &  -   &    -     &    -     &	  -  &	-	 & 17.3  \\
 5292.619  &                 &    0.648  &  -0.300   &   -    & 42.6 &  104.0  &  15.0 &   -  &    -     &  16.9   &	  -  &  21.4     & 22.8  \\
 5322.772  &                 &    0.482  &  -0.315   &    -   &   -   &  109.6  &  12.4 &  -   & 26.1    &    -     &	  54.5 &	-	 &      - \\
 5892.251  &                 &    1.439  &  -0.352   &    -   &   -   &  41.8  &  11.1 &  -   &    -     &    -     &	 -   &	-	 &      - \\
 6278.676  &                 &    1.196  &  -0.630   &    -   & 63.8 &         & -   &   -  &      -     &   -      &     -   &    -      &      -   \\
 4021.327  &  Nd II&   0.320  &  0.230    &       &   -   & 	-  &  61.8 &  -   &    -     &    -     &	  -  &	-	 &      - \\
 4059.951  &                  &    0.205 &   -0.360   &       &   -   &  60.3 &   -   &  -   &    -     &    -     &	  -  &	-	 & 31.2  \\
 4061.080  &                  &    0.471  &   0.550   &64.0  & 56.6 &  123.2  &  67.1 &   -  &    -     &    -     &	  -  &	-	 &      - \\
 4069.270  &                   &   0.060  &  -0.400   &   -    &   -   & 	  &   -   &  -   &    -     &    -     &	  -  &	-	 & 35.4  \\
 4109.448  &                   &   0.320  &  +0.180   &    -   &   -   &  111.4  &   -   &  -   &    -     &    -     &	  -  &	-	 &      - \\
 4446.390  &                  &   0.200  &  -0.630    &37.4  &  -     &  175.0  &  49.2 &60.4 &    -     &  32.1  &  	  -  &  47.1     & 44.5  \\
 4451.563 &                   &    0.381  &  -0.040   & -      &   -   & 	  &  74.1 &  -   &    -     &    -     & 104.9   &	-	 &      - \\
 4451.980 &                   &   0.000  &  -1.340    &  -     &   -   &  153.2  &  31.7 &  -   &    -     &    -     &	  -  &  36.4     & 44.6  \\
 4556.133  &                  &    0.064  &  -1.610   &  -     &   -   &  119.0  &   -   &  -   &    -     &    -     & 60.5   &-	         &      - \\
 4797.153 &                   &   0.559  &  -0.950    &  -     &   -   &  94.0  &  19.9 &  -   &    -     &    -     &	  -  &	-	 & 27.9  \\
 4811.342 &                   &    0.063  &  -1.140   &27.8  & 44.2 &  124.4  &  38.8 &   -  & 57.4    &  26.6  & 68.5   & 28.0      & 42.8     \\
 5089.832 &                   &    0.204  &  -1.160   &  -     &   -   & 	  &  14.9 &  -   &    -     &    -     &	 -   &	-	 & 16.7  \\
 5130.590 &                   &    1.300  &  +0.100   &  -     &   -   & 	  &  39.3 &  -   &    -     &    -     &	 -   &	-	 & 60.9  \\
 4859.039 &                    &   0.320  &  -0.830   &  -     &   -   &  147.6  &   -   &  -   &    -     &  44.5   &	 -   &	51.91    &      - \\
 4947.020 &                    &   0.559  &  -1.250   &  -     & -     &  68.1  &   -   &   -  &    -     &    -     &	 -   &	-	 &      - \\
 4961.387 &                    &   0.630  &  -0.710   &  -     & 19.9 &  113.3  &   -   &   -  & 42.3    &    -     &72.5    &	-	 &      - \\
 4989.953  &                    &  0.680   & -1.400   &  -     &   -   &  144.9  &   -   &  -   &    -     &    -     &	 -   &	-	 &      - \\
 5212.361 &                    &   0.204  &  -0.870   &  -     & 52.1 &  138.7  &   -   &   -  &    -     &    -     &	 -   & 22.7     & 36.2  \\
 5293.169 &                    &   0.823 &   -0.060   &39.0  &   -   & 	  &   -   &  -   &    -     &    -     &	 -   & 46.8     & 60.2  \\
 5276.878 &                    &   0.859 &   -0.440   &  -     &   -   &  93.0  &   -   &  -   &    -     &    -     &	 -   &	-	 &      - \\
 5287.133 &                    &   0.745  &  -1.300   &  -     & 14.9 &  71.6  &   -   &   -  &    -     &    -     &	 -   &	-        &      - \\
 5311.480 &                    &   0.990 &   -0.560   &  -     &   -   &  117.8  & 18.2 &  -   &     -     &    -     &47.8    & -         &      -  \\  
 5319.820  &                   &   0.550 &   -0.210   & 35.8 &   -   &      	 &  46.1 &  -   &     -     &    -     &76.9    & 51.5     & 59.3   \\ 
 5356.967   &                   &  1.264 &   -0.250   &  -     &   -   &  81.5 &   -   &  -   &    -     &    -     &	 -   &	-	 &      - \\
 5361.510  &                   &   0.68  &   -0.400   &  -     &   -   & 	  &  -   &  -   &     -     &  35.2 &61.4      &  49.9    &      -    \\
 5371.927  &                    &  1.412 &    0.003   &   -    &   -   & 	  &   -   &  -   &    -     &    -     &	 -   &	-	 & 30.1  \\
 5603.648 &                    &   0.380 &   -1.830   &   -    &   -   &  111.5  &   -   &  -   & 44.7    &  13.8   &	 -   &	-	 &      - \\
 5718.118 &                  &     1.410  &  -0.340   &   -    &   -   &  68.1  &   -   &  -   &    -     &    -     &	 -   &	-	 &      - \\
 5825.857 &                   &    1.081  &  -0.760   &   -    &   -   &  74.9  &   -   &  -   &    -     &    -     &	 -   &	-	 &      - \\
 4318.927  & Sm II&    0.280  &  -0.270   &   -   &    -   &  151.6  &  -    &37.7 &    -     &    -     &53.0    &     -     &      -   \\ 
 4424.337  &       &   0.485 &  -0.260    &     &   -    &        &   -    &  -    &    -     &    -     &   70.0  &	-	 &      - \\
 4434.318  &                   &    0.378 &   -0.576   & 53.5 & - &  116.9  &   -   &60.4 &    -     &  26.1   &	 -   &	-	 & 52.8  \\
 4499.475  &                   &   0.248 &   -1.413   &   -    &   -   &  71.9  &  10.8 &  -   &    -     &    -     &28.9    &-	         &      - \\
 4519.630 &                    &   0.540 &   -0.430   & 20.2 &   -   & 	 -&  28.3 &  -   &    -     &    -     &	 -   &	-	 & 37.5  \\
 4577.690 &                    &   0.250 &   -0.770   &   -   &       &  	- &   -   &22.7 & 36.4    &    -     &	 -   &	-	 &      - \\
 4566.210  &                   &   0.330  &  -1.245   & 13.8 &   -   & 	- &   -   &  -   &    -     &    -     &44.4    &-		 & 16.7 \\ 
 4674.600  &                   &   0.180 &   -0.560   &   -    &   -   & 	- &  27.4 &  -   &    -     &  27.9  &	 -   &	-	 & 37.7  \\
 4704.400  &                   &   0.000 &   -1.562   & 30.2 &   -   & 	- &   -   &  33.1 &    -     &    -     &	 -   & 23.8     & 32.8  \\
 4615.444  &                  &    0.544 &   -1.262   &   -    &   -   &       - &   -   &  -   &    -     &    -     & 18.7   &	        &      - \\ 
 4726.026  &                  &    0.333 &   -1.849   &   -    &   -   &        -&   -   &  -   &    -     &    -     & 29.8   &	-        &      - \\
 4791.580  &                   &   0.104 &   -1.846   &   -    &   -   &  77.5  &   -   &  -   &    -     &    -     & 24.0   &	-        &      - \\
 4815.805  &                   &   0.185 &   -1.501   &   -    &   -   &         &   -   &  -   &    -   - &    -     &	-    & 16.2     &      -  \\
 4129.700  & Eu II  &   0.000 &   +0.204   &       &   -   & 	 -&   74.9  &  -   &    -     &    -     &	-    &  -    & 80.3  \\
 4205.050  &                 &    0.000  &  +0.117   &    -   &   -   & 	 -&   -   &  -   &    -     &    -     &	-    &	-	 & 97.2  \\
 6437.640  &                  &   1.319 &   -0.276   &    -   &14.8  &  27.3  &   -   &   -  &    -     &    -     &	-    &	-	 &      - \\
 6645.130  &                  &   1.380  &  +0.204   &    -   &   -   &  61.9  &   -   &  -   &    -     &    -     &	-    &	-	 &      - \\
 4103.310  &  Dy II  &   0.103 &   -0.346   &       &     -  &         &   -   &68.9 &    -   - &    -     &   46.7 & -         &      -   \\ 
 4923.167  &         &   0.103  &  -2.384   &       & 26.1 &         & -     &   -  &    -   - &    -     &      -  & -         &      -  \\

 \hline
\end{tabular}
\end{table*}

\section{Discussion on individual stars}
Comparisons of our estimated atmospheric parameters and elemental 
abundance ratios with literature values whenever available, are presented in 
Tables 10 and 11 respectively. In the case of HD~201626 the author gave 
absolute abundances (log$\epsilon$(x), rather than abundance ratios; 
in Table 11, calculated abundance ratios are presented using solar 
Fe values from Asplund et al. (2005).

Information on the circumstellar environment of these objects are not 
available in the literature. Information on polarization estimates is 
limited to only two objects in this sample, i.e. HD~81192 and 125079. 
Estimated percentage polarization in B, V, R, I bands are found to be low
with $<$ 0.2 per cent in all the four bands for HD~81192 and $<$0.4 per cent for HD~125079
(Goswami \& Karinkuzhi 2013). \\

$\bf{HD~125079, 4395, 216219}$: These three low-velocity CH stars 
are identified as CH sub-giants by Bond (1974). Spectral characteristics 
of sub-giant CH stars are similar to giant CH stars. With lower 
temperatures  and luminosities the CH sub-giants occupy a position 
below the giants in H-R diagram. High resolution spectroscopic analyses 
of sub-giant CH stars   by Sneden $\&$ Bond  (1976) and  Luck $\&$ 
Bond (1982) have confirmed in this type of objects the enhancement 
of heavy elements  and  a metal deficiency  in the range -0.2 to -0.5. 
As in the case of CH giants, abundance peculiarities of sub-giant 
CH stars are also explained  as due to mass transfer mechanisms 
from their binary companions (Luck \& Bond 1991). 
Smith et al. (1993) have  reported  the abundances for 
elements Y, Zr, Ba and Nd. The effective temperature adopted by us is 
about  250 K larger than what they have considered for these three stars. 
The estimated  metallicity  values are about 0.15 dex lower than their
estimates. For HD~125079, our estimated  abundances for 
Y is about 0.3 dex higher and for  Ba and Nd, the values are 
about 0.2 dex higher. In addition, we have also estimated the abundances 
of Ce, Pr and Sm in these objects. For the other two stars we have 
derived almost similar  abundances for Y. For Zr, Ba and Nd, our 
estimated abundances  are slightly higher than their values. 

$\bf{HD~16458}$: This star is included both in the Ba star catalogue 
of L\"u (1991) and CH star cataloge of Bartkevicius (1996). Estimated 
radial velocity is about 18.24 km s$^{-1}$. Chemical
composition studies of this object with respect to the abundances of
a standard giant star $\beta$ Gem by Tomkin $\&$ lambert (1983) have 
suggested  enhancement of heavy elements in this object. Smith (1984)
also studied this star and reported the enhancement of heavy elements.
Our abundance estimates  are slightly higher compared to Smith (1984). 
With  [Ba/Fe] 
value ${\sim}$ 1.18 and [Ba/Eu] ${\sim}$ 0.50 this star satisfies the 
conditions for  CEMP-r/s stars (Beers $\&$ Christlieb 2005). Long-term 
radial velocity monitoring of HD~16458 by McClure $\&$ Woodsworth (1990) 
have confirmed it to be a binary star.

$\bf {HD~5395}$: HD 5395 is also included both in the Ba star catalogue 
of L\"u (1991) and CH star cataloge of Bartkevicius (1996). 
 Mcwilliam (1990) provided the elemental abundances for Sr, Zr, Y, La 
and Nd. We present updates on these abundances along with the first 
estimates of abundances for Ba, Ce, Pr and Eu.  While Y, Ba and Ce 
show almost near-solar values, Sr and La show  mild enhancement with 
[Sr/Fe]=0.26 and [La/Fe]=0.24. Pr, Nd and  Eu  show larger enhancement 
with respect to Fe.  As far as the chemical abundances are concerned this
object does not seem to belong to the group of CH stars if we stick to the
definition that  CH stars are those that  show  high abundance of s-process 
elements with [Ba/Fe] $>$ 1.

$\bf{HD~214714 and~188650}$: The spectra of  HD 214714 and  188650 
show a close resemblance. Bidelman (1957) first noticed strong bands of 
CH and moderate CN bands in the spectra of these stars and called them as 
low-velocity CH stars. Our estimated radial velocities for these two 
objects are $-$7.04 and $-$24.6 km s$^{-1}$ respectively. Strong ionised 
lines of elements seen in their spectra indicate high luminosity. However, 
following Morgan, Keenan $\&$ Kellman (1943), Luck (1991)  referred to these
objects   as a cyanogen weak giants as their spectrum is characterised by 
weak violet CN bands. Using a curve of growth method in comparison 
with $\beta$ Aqr, Baird et al. (1975) estimated chemical abundances 
for these objects and reported  a high lithium abundance for HD 214714. 
Our chemical analysis   with respect to solar values indicates  a mild enhancement 
of heavy elements. We present first-time estimates of chemical abundances 
for HD~188650 based on  high-resolution spectrum. The abundance of Eu could 
not be estimated for this object.
Light s-process element Y, and heavy s-process elements 
Ba, Ce and Sm show near-solar values in this object. 
Pr and Nd also show mild enhancement with  [Pr/Fe] ${\sim}$ 0.57 and 
[Nd/Fe] ${\sim}$ 0.39 respectively. This object does not seem to 
represent  the characteristic properties of CH stars as far as the 
heavy element abundances are concerned.

$\bf{HD~201626}$ : This object was first   identified as a CH star by 
Wallerstein $\&$ Greenstein (1964);  its spectrum closely resembles 
the spectrum of the  well-known CH star HD~26. 
This is a high velocity object with a radial velocity of
$-$141.6 km s$^{-1}$ and  metallicity  [Fe/H] ${\sim}$ $-$1.4. 
Abundances of  heavy elements Ba, La, Ce, Pr, Nd, Sm and Dy  with
respect to Fe  are  found to be highly enhanced in this object. 
Our estimates of heavy element abundances  are in good agreement 
with the  estimates  of Vanture (1992c)  

$\bf{HD~48565}$: The object HD~48565  is known to  show abnormally strong 
Sr line at 4077 {\rm \AA}  (North et al. 1994). Bidelmann (1981) 
classified these  type of objects  as F Str ${\lambda}$ 4077 stars. 
They exhibit enhancement of light and heavy s-process elements but  
abundances of iron-peak elements are similar to those generally seen 
in F type stars. Allen \& Barbuy (2006a) have given the abundance 
estimates for heavy elements in this star. Our results are slightly 
higher than  their results, with  [X/Fe] values $>$ 1 for all the 
heavy elements (Table 11). Our estimated radial velocity 
($-$25.74 km s$^{-1}$) is about 6 km/s lower than the literature value. 
This object is showing a large radial velocity variations 
(North et al. (1994), Nordstr\"oem et al. (2004)) giving 
indications of the object being a  binary system.

{\footnotesize
\begin{table*}
{\bf Table 10. Atmospheric parameters from  literature }\\

\begin{tabular}{lccccc}
\hline

Star name &  Vmag  &  Teff  &  log g & [Fe/H]  &  Reference\\
\hline
HD4395    &  7.50 	& 5550   &  3.66  & -0.17   &1\\
          &     	& 5478   &  3.40  & -0.31   &8\\
          &     	& 5478   &  3.30  & -0.38   & 2\\
          &             & 5450   &  3.30  & -0.33   & 3\\
          &             & 5467   &  3.32  & -0.35   &9 \\
          &             &        &        &         &   \\
HD5395    &  4.63 	& 4860 	 &  2.51  & -0.24   &1\\
	  &             & 4800   &  2.20  & -1.00    &6 \\
	  & 	       	& 4770   &  2.90  & -0.51        &11 \\
          &             & 4764   &  2.36  & -0.44   & 9\\
          &             &        &        &         &   \\
HD16458	  &   5.78 	&4550    & 1.8	  & -0.65   & 1\\
          & 	        &4582    & 2.00	  &-0.32    &4 \\
          &	        &4582    & 2.00	  &-0.43    &4 \\
          & 	        &4800    & 1.80	  &-0.30    &5\\ 
          & 	        &4500    & 1.40  &-0.36    &6 \\
          &             &        &        &         &   \\
HD48565	  &   7.07 	& 6030   & 3.80	  & -0.59   & 1\\
	  &            	& 5910 	 & 4.27	  & -0.56   & 8\\
          &             &        &        &         &   \\
HD 81192  &   6.54      & 4870   & 2.75	  & -0.50   &1 \\
          & 	       	& 	 & 	  & -0.61   &12 \\
          & 	       	&  4755  & 2.40	  & -0.60   &13 \\
          & 	       	& 4582   & 2.75	  & -0.70   &14 \\
          &             &        &        &         &   \\
HD125079  &  8.60 	& 5520   & 3.30	  & -0.18    & 1\\
          &             & 5305   & 3.50	  & -0.30    & 2\\ 
          &	        & 5300   & 3.50   & -0.16    &3 \\
          &             &        &        &         &   \\
HD188650  &   5.76      &  5700  & 2.15   & -0.46   &1\\
          &             &        &        &         &   \\
HD214714  &   6.03      &  5550  & 2.41	  & -0.36  &1 \\
          &             &  5400  & 2.38   & -0.36  & 7\\
          &             &        &        &         &   \\
HD216219  & 	 7.40   &  5950  & 3.5	  & -0.17  & 1\\
          & 	 	&  5478  & 2.80	  & -0.55  & 8\\
          &  	   	&  5600  & 3.25   & -0.39  &2 \\
          &  	   	&  5600  & 3.20   & -0.32  &3\\
          &             &        &        &        &   \\
HD 201626 &     8.13    & 5120   &        &-1.39  &1 \\
          &             &        &        &        &   \\
\hline
\end{tabular}

{\bf References.}1. Our work 2. Smith $\&$ Lambert  1986, 3. Smith et al. 1993,			    
4. Tomkin $\&$ Lambert 1983, 5. Smith 1984,  \\6. Villacanas et al. 1990,   7. Luck  1991,  8. Krishnaswamy $\& $Sneden  1985, 9. Soubiran et al. 2008,  10. North et al. 1994,\\
11. McWilliam  1990,  12. Luck $\&$  Bond 1983,
13. Luck $\&$ Bond  1985, 14. Cottrell $\&$ Sneden 1986 \\ 
\end{table*}
}
{\footnotesize
\begin{table*}\tiny
{\bf  Table 11. Elemental Abundances from literature}\\
\begin{tabular}{lllllllllllll}
\hline
Star Name& [Sr I/Fe]&[Y II/Fe]&[Zr II/Fe]&[Ba II/Fe]&[La II/Fe]&[Ce II/Fe]&[Pr II/Fe]&[Nd II/Fe]&[Sm II/Fe]&[Eu II/Fe]&[Dy II/Fe]& reference\\
\hline
Subgiant-CH stars\\
HD 4395 &1.08 & 0.65& 0.58& 0.79& 1.03&  0.42  &0.53  &0.80 & 1.08&  -&-&1\\
        &-    &0.70& 0.61& 0.53& - & -&-  &0.39 & -&  -&-&2\\
        &1.08 &0.10& 1.01& -0.20& - & -  &-  &-0.20 & -&  -0.55&-&3\\
HD 125079& 1.59 &1.05& - &1.06&    - &0.93 & 1.00& 1.16& 0.56 &-&-&1\\
         & - &1.36& 0.98 &0.89&    - &- & -& 0.99& - &-&-&2\\
HD 216219 &1.80 & 1.00 &0.98 &1.10 & 1.04& 1.03& 1.14 &0.99 &0.91 & 0.07&-&1\\
          &- & 1.10 &0.85 &0.87 & -& 0.95& -&-&- & -&-&2\\
          &1.30 & 0.96 &- &- & 0.72& 0.70& -&0.83&0.53 & -&-&3\\
\hline
CH stars\\
HD 5395 &0.26& 0.05& - &0.03 & 0.24 & 0.06& 0.79& 0.74 &-& 0.34&1.38&1\\
        &1.44& 0.27& -1.04 &- & 0.17 & -& -& 0.38 &-& -&-&4\\
HD 16458 & 1.37 &1.46& 1.17& 1.18 &1.42& 1.47& 1.82 &1.55& 1.87& 0.66 &-&1\\
         & 1.25 &1.06& 0.95& 1.03 &0.96& 0.89& 0.68 &0.97& -& 0.35 &-&5\\
HD 48565& 1.73& 1.08 & 0.9 & 1.52 &1.46 &1.42 &1.29& 1.51 &1.18 & 0.29&-&1\\ 
        & 0.97& 1.01 & 1.19 & 1.29 &1.39 &1.60 &1.09& 1.31 &0.95 & 0.35&-&6\\
HD 81192 &0.58& 0.10 &0.12 & 0.13 &  -0.13 &-0.15 &- &1.01 & 0.85 &-&1.21&1\\
         & -&-0.07 &-0.34 & -0.12 &  -0.51 &-0.04 &- &- & - &-&-&7\\
HD 188650&- &-0.03 &-& -0.01 &- &-0.03& 0.57& 0.39 &-0.12& -&-&1\\
HD 214714 &-&0.22 & -0.28 &-0.31& -& 0.05& 0.93 &0.36& 0.46 &-&-&1\\
          &-&0.24 & 0.72 &-& -& 0.36& - &-& - &-&-&8\\
HD 201626& - &- &- &2.12& 1.76& 1.89& 2.09 &2.24& 1.63& -&0.97&1\\
         & - &0.09 &- &-& 1.6& 1.62& 2.09 &1.55&1.6 & -&&9\\
\hline
\end{tabular}
{\bf References.}1. Our work 2.Smith et al. (1993) 3. Luck \& Bond (1982) 4. McWilliam (1990) 5. Smith (1984) 6. Allen \& Barbuy (2006a) \\
7. Luck \& Bond (1985) 8. Luck (1990)  
9. Vanture (1992c)
\end{table*}
}

$\bf{HD~81192}$: Except for Sr, Nd and Sm this object shows almost
near-solar values for Y, Zr, Ba, La, and Ce with respect to Fe.
Morgan, Keenan $\&$ Kellman (1943) have noted weaker CN bands 
in HD 81192 compared to other stars of same temperature and luminosity. 
Weakening of CN band is most common in stars with high space velocities. 
{\bf Estimated} radial velocity of this object is 136 km s$^{-1}$.  
Cottrell et al. (1986) have studied the kinematics and elemental 
abundances of this object. Estimated heavy element abundances  by  
Luck $\&$ Bond (1985) are found to be in close agreement with our 
estimates. We present first-time estimates of  abundances for Sr, Sm 
and Dy for this object. With [Ba/Fe] = 0.13, this object too 
does not seem to belong to the group of CH stars.

\subsection{Stellar masses}
We derived the  mass of the program stars from their locations in
the Hertzsprung-Russel diagram  (Figures 8 to 10), using  Girardi 
et al. (2000) database of evolutionary tracks in the mass range of 
0.15 M$_{\odot}$ to 7.0 M$_{\odot}$ and the Z values from 0.0004 to 0.03.  
These evolutionary tracks are available at http://pleiadi.pd.astro.it/. 
 Since [Fe/H] of our target stars are 
near solar, we have selected an initial composition of Z=0.0198, Y=0.273. 
 The masses derived using spectroscopic temperature estimates are presented  
in Table 12.  For four stars in our sample 
that have metallicities $<$ $-$0.5 we also used evolutionary tracks 
for Z = 0.008, but  the masses obtained are found to be  similar to those 
obtained using evolutionary tracks with Z = 0.019. Derived stellar
masses are in general $<$ 2 M$_{\odot}$, except for HD~188650 and 214714,
for which our estimated stellar masses are  3.5 and 4 M$_{\odot}$ respectively.

\begin{figure}
\centering
\includegraphics[height=8cm,width=8cm]{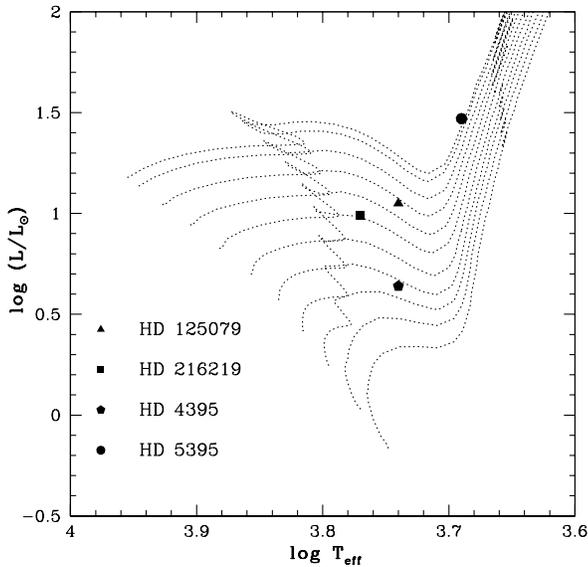}
\caption{The location of HD~125079, HD~216219, HD~4395 and HD~5395 are
indicated  in the H-R diagram. The masses are
derived using the evolutionary tracks of Girardi et al. (2000).
 The evolutionary tracks for masses 1, 1.1, 1.2, 1.3 1.4, 1.5, 
1.6 1.7, 1.8, 1.9 and  1.95 M$_{\odot}$ from bottom to top are shown 
in the  Figure.}
\end{figure}

\begin{figure}
\centering
\includegraphics[height=8cm,width=8cm]{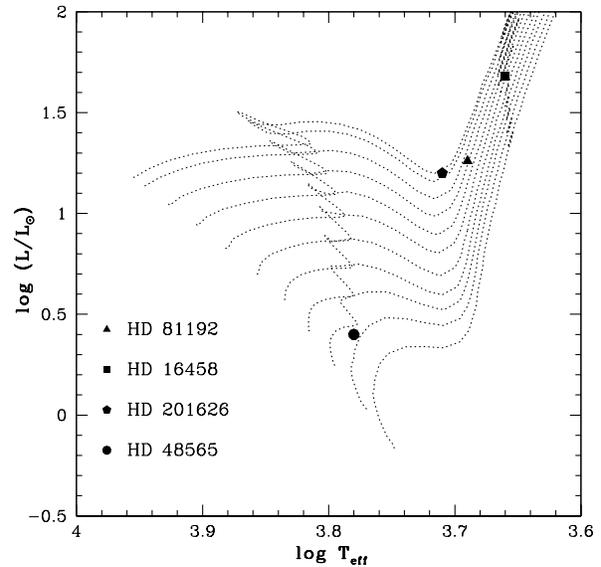}
\caption{
The location of HD~81192, HD~16458, HD~201626 and HD~48565 are 
indicated  in the H-R diagram. The masses are
derived using the evolutionary tracks of Girardi et al. (2000).
The evolutionary tracks are shown for masses 1, 1.1, 1.2, 1.3 1.4, 1.5, 
1.6 1.7, 1.8, 1.9 and  1.95 M$_{\odot}$ from bottom to top.}
\end{figure}

\begin{figure}
\centering
\includegraphics[height=8cm,width=8cm]{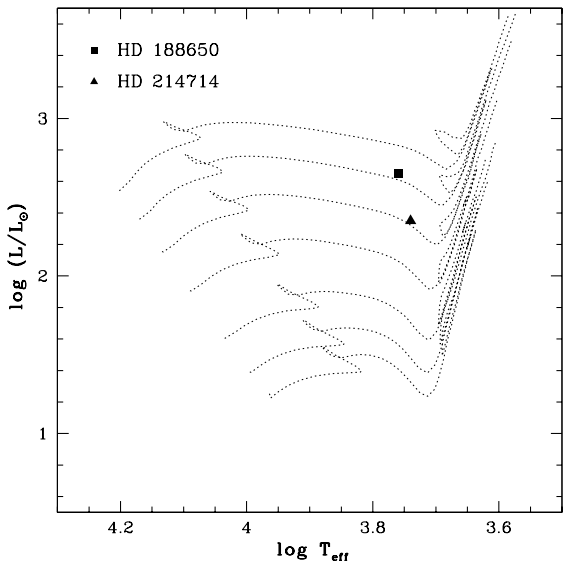}
\caption{
The location of HD~188650, and  HD~214714 are indicated  in the 
H-R diagram. The evolutionary tracks of Girardi et al. (2000) 
are shown for masses  2, 2.2, 2.5, 3.0, 3.5, 4.0 and 4.5 M$_{\odot}$ 
from bottom to top.}
\end{figure}

\begin{table*}
{\bf  Table 12: Stellar Masses }\\
\begin{tabular}{lccc}
\hline
Star Name.   &$M_v$& $log(L/L_{\odot})$ & $Mass(M_{\odot})$  \\
\hline
HD~4395    &	2.8&	0.64&1.35\\
HD~5395    &	0.7&	1.47&1.95\\
HD~16458   &	0.1&	1.68& 1.5\\
HD~48565   &	3.7&	0.4&1.15\\
HD~81192   &	1.2&	1.26&1.7\\
HD~125079  &	1.8&	1.05&  1.75\\
HD~188650  &	-2.7&	2.65&4.0\\
HD~201626  &	1.3	&1.2&1.9\\
HD~214714  &	-1.3&	2.35&3.5\\
HD~216219  &	2.1&	0.99& 1.6\\
\hline-
\end{tabular}
\end{table*}

\section{Parametric model based study}
Elements heavier than iron are  produced mainly by 
two neutron-capture processes, the s-process and the r-process. 
Observed abundances of heavy elements estimated using model atmospheres and
spectral synthesis techniques do not provide direct quantitative estimates 
of the relative contributions from s- and/or r- process nucleosynthesis. 
We investigated ways to delineate the observed abundances into  
their respective r- and s-process contributions in the framework of a 
parametric model using an appropriate model function.
The origin of the n-capture elements   is explored by comparing the 
observed abundances with predicted s- and r- process contributions 
following Goswami et al. (2010a, and references there in). Identification 
of the dominant processes contributing to the heavy element abundances 
in CEMP stars is likely to provide clues to the origin of the  
observed abundances. The ith element abundance can be calculated as 
                        
         $N_i$(Z)  = $A_sN_{is}$  + $ A_r N_{ir}$  10$^{[Fe/H]}$

where  Z is  the metallicity of the star, $N_{is}$  indicates the  abundance  
from s-process in  AGB  star,   $N_{ir}$ indicates the  abundance  
from r-process; $A_s$ indicates  the component coefficient that 
correspond to   contributions   from  the s-process  and
$A_r$ indicates  the component coefficient that correspond to contributions 
  from the r-process.
 
We utilized the solar system s- and r-process isotopic abundances 
from stellar models of  Arlandini et al. (1999). The observed elemental 
abundances were scaled to the metallicity of the corresponding CH star 
and normalised to their respective Ba abundances. Elemental abundances 
were then fitted with the parametric model function. The best fit coefficients 
and reduced chi-square values for a set of CH stars are given in Table 13. 
 The best fits obtained with the parametric model 
function $log{\epsilon_i}$  = $A_sN_{is}$  + $ A_r N_{ir}$ for HD~16458,
48565, 125079, and 216219  are shown in Figures 11 -14. 
The errors in the derived abundances play an important role in deciding 
the goodness of fit of the  parametric model functions. 
From the parametric model based analysis we find  HD 16458 to fall 
in  the group of CEMP r/s stars and the  stars  HD~48565, HD~125079, and 
HD~216219 belong to the group of CEMP-s stars.  
                                       
\begin{table*}
{\bf  Table 13: Best fit coefficients and reduced chi-square values}\\
\begin{tabular}{lccc}
\hline
Star Name& $A_s $& $A_r$&$ {\chi}^2$\\
\hline
HD 16458 &0.49$\pm$ 0.09 & 0.60 $\pm$ 0.09 &  1.6\\
HD 48565 &0.835$\pm$ 0.09 & 0.112 $\pm$ 0.08 &  1.15\\
HD 125079  &0.832$\pm$ 0.16 & 0.182 $\pm$ 0.15 &  0.50\\
HD 216219&0.859$\pm$ 0.14 & 0.169 $\pm$ 0.13 &  1.22\\
\hline
\end{tabular}
\end{table*}

\begin{figure}
\centering
\includegraphics[height=8cm,width=8cm]{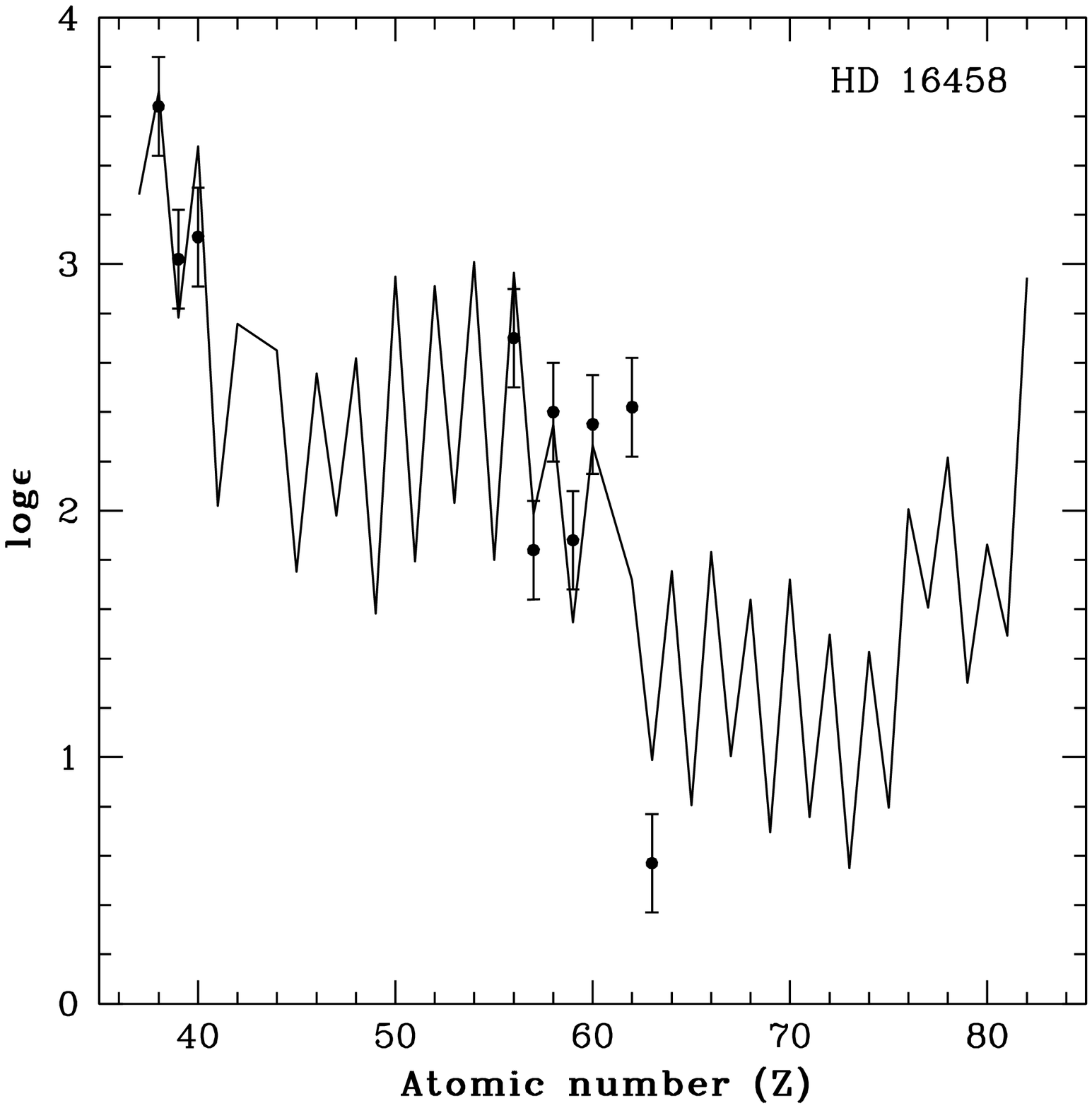}
\caption{  Solid curve represent the best fit for the parametric model
function log${\epsilon}$  = A$_{s}$N$_{si}$ + A$_{r}$ N$_{ri}$, where N$_{si}$
 and N$_{ri}$ represent the abundances
due to s- and r-process respectively (Arlandini et al. 1999, Stellar model, scaled
to the metallicity of the star). 
The points with errorbars
 indicate the observed abundances in  HD~16458.}
\end{figure}

\begin{figure}
\centering
\includegraphics[height=8cm,width=8cm]{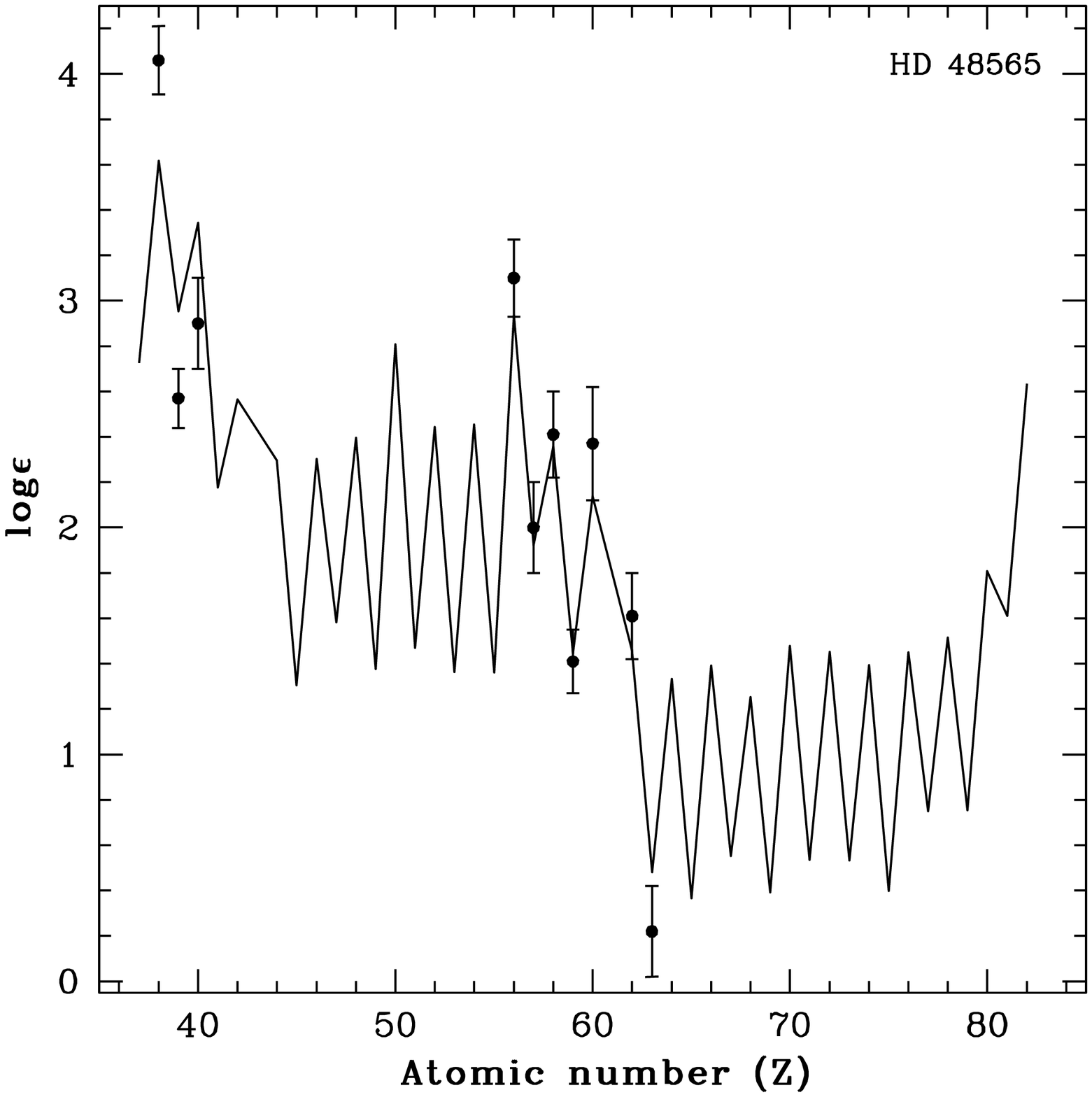}
\caption{  Solid curve represent the best fit for the parametric model
function log${\epsilon}$  = A$_{s}$N$_{si}$ + A$_{r}$ N$_{ri}$, where N$_{si}$
 and N$_{ri}$ represent the abundances
due to s- and r-process respectively (Arlandini et al. 1999, Stellar model, scaled
to the metallicity of the star). 
The points with errorbars
 indicate the observed abundances in  HD~48565.}
\end{figure}

\begin{figure}
\centering
\includegraphics[height=8cm,width=8cm]{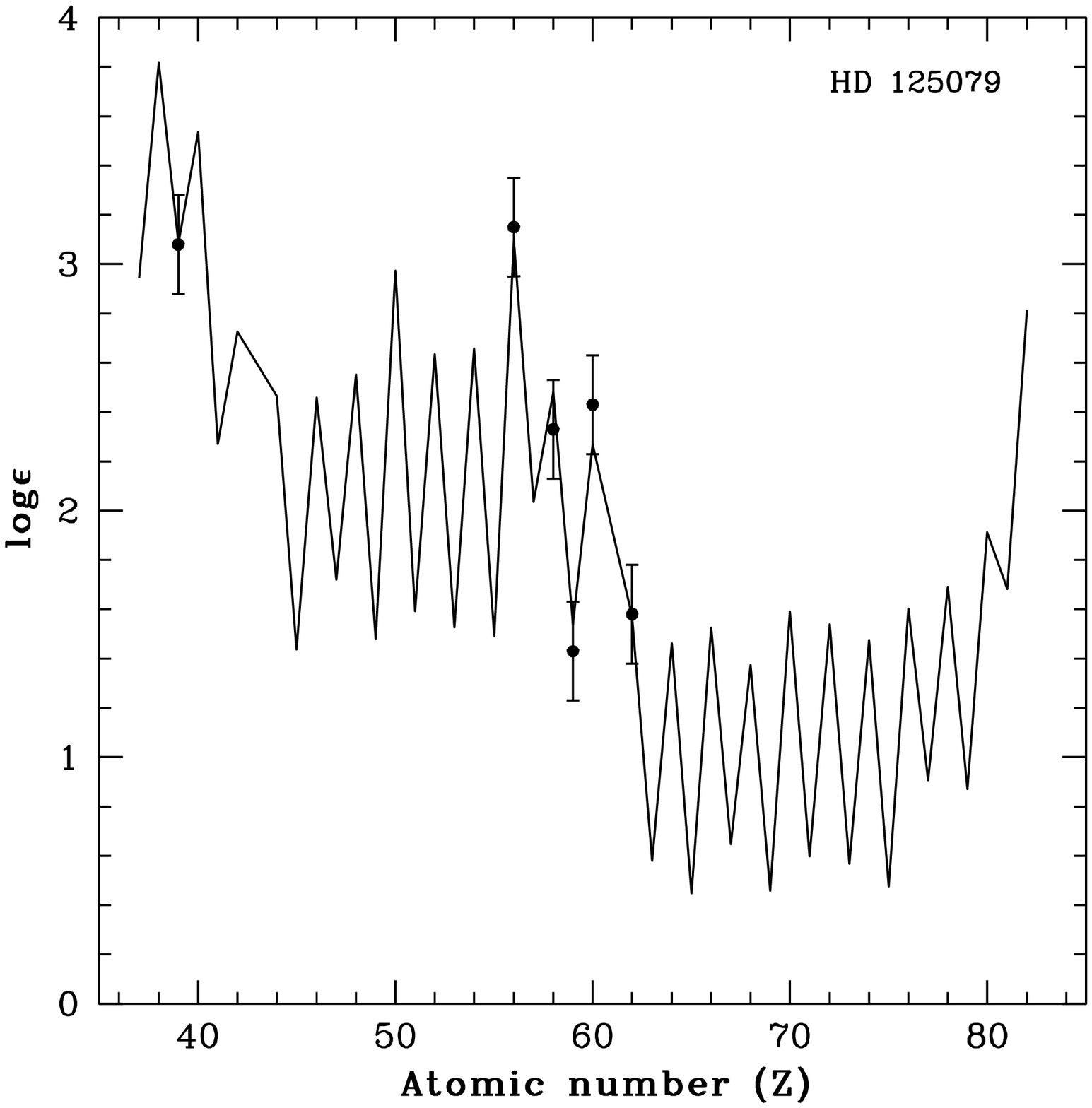}
\caption{  Solid curve represent the best fit for the parametric model
function log${\epsilon}$  = A$_{s}$N$_{si}$ + A$_{r}$ N$_{ri}$, where N$_{si}$
 and N$_{ri}$ represent the abundances
due to s- and r-process respectively (Arlandini et al. 1999, Stellar model, scaled
to the metallicity of the star). 
The points with errorbars
 indicate the observed abundances in  HD~125079.}
\end{figure}

\begin{figure}
\centering
\includegraphics[height=8cm,width=8cm]{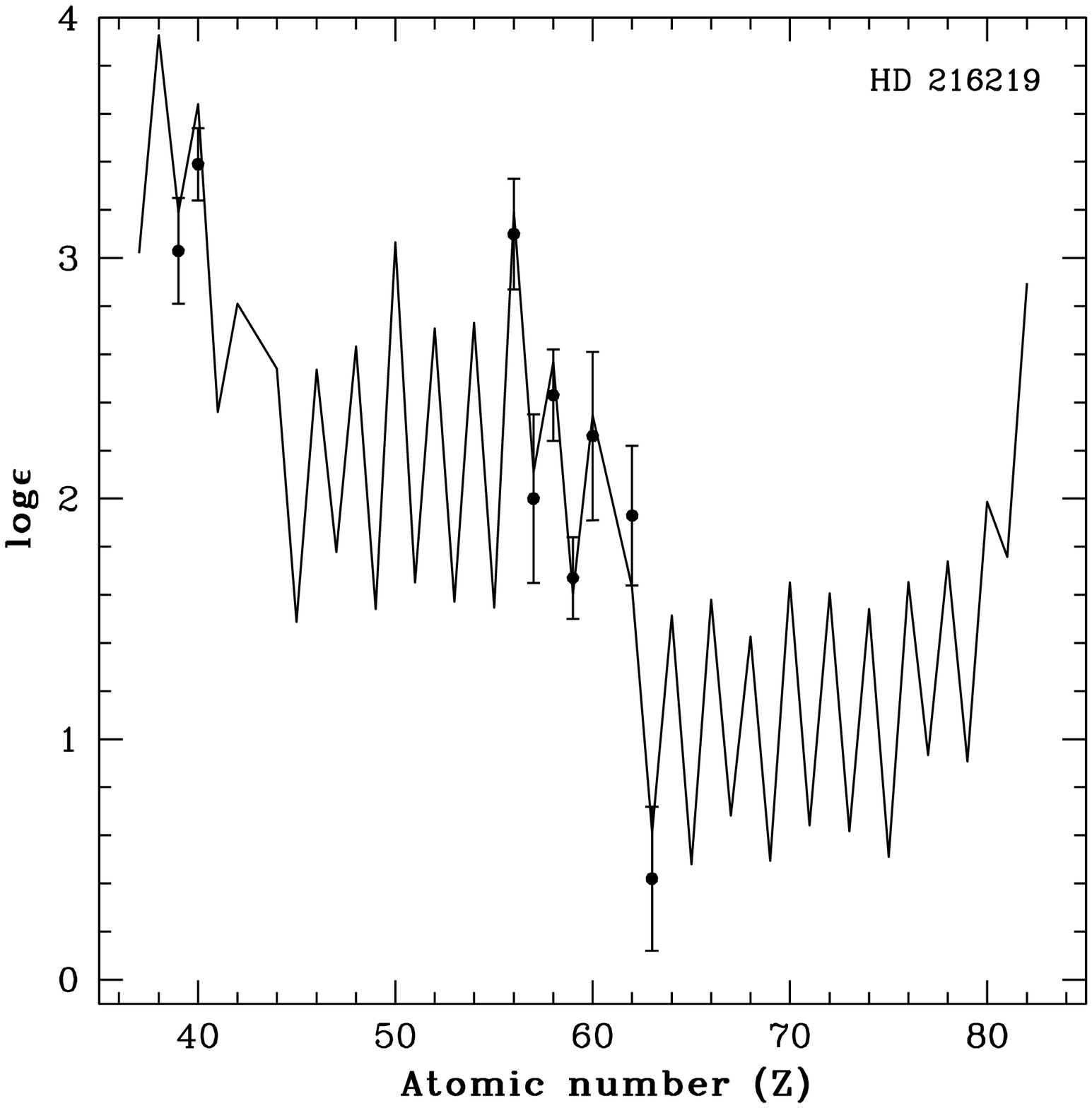}
\caption{  Solid curve represent the best fit for the parametric model
function log${\epsilon}$  = A$_{s}$N$_{si}$ + A$_{r}$ N$_{ri}$, where N$_{si}$
 and N$_{ri}$ represent the abundances
due to s- and r-process respectively (Arlandini et al. 1999, Stellar model, scaled
to the metallicity of the star). 
The points with errorbars indicate the observed abundances in  HD~216219.}
\end{figure}

\section{Conclusions}
Elemental abundances are presented for twenty two elements for a set of
ten stars  from the CH star catalogue of Bartkevicius (1996). The sample
includes  eight low-velocity CH 
stars (V$_{r}$ ${\le}$ $\pm$ 50 km s$^{-1}$) and two high-velocity
(V$_{r}$ ${\ge}$ $\pm$ 100 km s$^{-1}$) CH stars. Metallicity [Fe/H] of 
the two high-velocity
stars HD~201626 and 81192 are found to be  $-$1.4 and $-$0.5 respectively.
The low-velocity stars metallicity ranges from $-$0.18 to $-$0.66.
While Ba abundance is measured in all the ten stars, we could measure 
the Eu abundance only for four stars. 

Beers \& Christlieb (2005) have used the Ba and the Eu abundances
 to classify CEMP  stars into several broad  categories, among which
CEMP-s stars are those  that exhibit large over-abundances of
s-process elements with [Ba/Fe] $> $+1 and [Ba/Eu] $>$ +0.5. In this
classification scheme  CEMP-r/s, stars are those with
0 $<$ [Ba/Eu] $<$ +0.5; they   exhibit both r- and s-process enhancements.
CEMP-s stars are believed to be  the metal-poor counterparts of CH stars
having  same origin as CH stars.  The  observed  chemical 
composition of CEMP-s stars is also explained considering  a binary picture,
as in the case of CH stars. 
Two subgiant CH stars and three objects from the CH star catalogue, i.e.
HD 16458 ([Ba/Fe]=1.18, [Ba/Eu]=0.52), HD 48565 ([Ba/Fe]=1.52, [Ba/Eu]=1.23),
HD 201626 ([Ba/Fe]=2.12),
satisfy this criterion for CEMP-s stars; among these three the first two
are also listed in the Ba star catalogue. CH stars and Ba stars are known to show
enhanced abundances of carbon and heavy elements. Barium stars are generally
believed to be  the metal-rich population I analogues  of CH stars.
Allen \& Barbuy (2006a,b) have concluded that barium stars also have same
s-process signatures of AGB stars similar to CH stars. Peculiar abundances
of heavy elements observed in barium stars are believed to be the result
of a mass transfer process and can be explained with the help of a binary
picture including low-mass AGB stars (Jorisson \&  Van Eck (2000)).

The relationship between CEMP-s and CEMP-r/s stars are not clearly 
understood; there are however speculations that the progenitors
of the CEMP-s and CEMP-r/s class may be one and the same (TP-AGB)
(Tsangarides 2005). None of the four stars for which we could measure 
both Ba and  Eu abundances are found to satisfy the criterion of 
Beers \& Christlieb (2005) for CEMP-r/s stars.

Three of our program stars HD~16458, 201626, and 216219  with 
[Ba/Fe] $>$ 1,  are  known to be confirmed  binaries with  periods
 2018 days, 1465 days and  3871 days  respectively 
( McClure (1984, 1997), McClure \& Woodsworth 1990).
Long-term radial velocity monitoring of 10 years for the sub-giant CH star 
HD~4395 shows a radial velocity variation of $-$4 km s$^{-1}$ indicating
its binarity  (McClure 1983, 1984, 1997). 
Our radial velocity estimate differs by 6 km s$^{-1}$ from the literature
value. This object however gives [Ba/Fe] = 0.79 and does not satisfy
Beers \& Christlieb (2005) criterion for CEMP-s stars.

The chemical composition of HD~81192 with [Ba/Fe]= 0.13, is peculiar,
(i.e., it is enriched) in heavy elements of Nd and Sm and shows  near-solar 
abundances for Ba, La, and Ce. This object shows a mild enhancement of  
Sr  with [Sr/Fe] = 0.58. For CH and   CEMP-s  stars estimated [hs/Fe] 
are in general ${\ge}$ 1, where hs represents the heavy s-process elements 
and ls represents the light s-process elements. This condition is also 
not satisfied by  this object. The binary status of this object is not 
known; the object does not seem to represent a typical  CH star as 
far as its chemical composition is concerned.   

 HD~5395, with estimated  [Ba/Fe] ${\sim}$ 0.13,  [Eu/Fe] ${\sim}$ 0.34 
and [Ba/Eu] ${\sim}$ $-$0.31  also does not seem to belong either to 
the group of CH stars or CEMP-r/s stars. However, heavy elements Pr and Nd
are found to be overabundant with  [Pr/Fe] and [Nd/Fe] values of 0.79 
and 0.74 respectively.   Sr and La are also found to be  mildly 
overabundant  with [Sr/Fe] ${\sim}$ 0.26 and [La/Fe] ${\sim}$ 0.24. 
The abundance of HD~5395 is however consistent with one of the 
characteristic properties of CH stars, i.e.  the  2nd peak s-process 
elements are more abundant than the first-peak s-process elements.

Estimated values of [Ba/Fe] for the objects HD~188650  and 214714 are 
respectively  ${\sim}$ $-$0.01 and $-$0.31. Abundance of Eu could not 
be estimated for these two objects from our spectra. Except for Pr 
([Pr/Fe] = 0.57) and Nd ([Nd/Fe] = 0.39) all other heavy elements 
i.e., Y, Ce and Sm show near-solar values for HD~188650. Similarly, 
HD~214714 shows a near-solar value for Ce. While Y is overabundant 
with [Y/Fe] = 0.22, Zr is found to be underabundant with [Zr/Fe] = $-$0.28 
in this object. Pr, Nd and Sm are overabundant with [Pr/Fe] = 0.93, 
[Nd/Fe] = 0.36 and [Sm/Fe] = 0.46. It is possible that the objects that  
show mild enhancement of heavy elements such as Pr, Nd etc., their origin
could be from material that are  pre-enriched with such heavy elements.
CH stars are known as low-mass high velocity objects, however our estimated stellar
masses for HD~188650 and HD~214714 are  much higher than solar values with
4.0 M${_\odot}$ and 3.5 M${_\odot}$ respectively (Table 12).

Several authors have used [hs/ls] as a good indicator of s-process 
efficiency and used these values for classification of CH stars. For example
Bisterzo et al. (2012) classified the stars with [hs/Fe] value 
$\geq$ 1.5 as S II stars and those with [hs/Fe] value $\le$ 1.5 as 
S I stars. In our sample three objects belong to S II category 
according to these criteria.

Our  parametric model based study also  gives  higher values for the 
component coefficients corresponding  to contributions coming from 
the s-process than those  from the r-process (Table 13) indicating  
that the s-process is the dominant one for the production of heavy 
element in the objects HD~48565, 125079, and 216219.\\
 
{\it Acknowledgement}\\
We thank the referee B. Barbuy, for her valuable suggestions which have improved the paper considerably. This work made use of the SIMBAD astronomical database, operated at CDS, 
Strasbourg, France, and the NASA ADS, USA. DK was a junior research fellow with the DST 
project  NO. SR/S2/HEP-09/2007 during the early part of this work and   
currently a CSIR-senior research fellow. Funding from  DST  and  CSIR are gratefully 
acknowledged.\\

\end{document}